\documentclass[journal]{IEEEtran}
\usepackage[utf8]{inputenc}
\usepackage{graphicx}
\usepackage{mathtools}
\usepackage{amsmath}
\usepackage{float}
\usepackage{comment}
\usepackage{geometry}
\usepackage{caption}
\usepackage{booktabs}
\usepackage{xcolor}
\usepackage{mathrsfs}
\usepackage[]{hyperref}
\usepackage{cite}
\usepackage{array}
\usepackage{xcolor}
\usepackage{soul}

\ifCLASSOPTIONcompsoc
  \usepackage[caption=false,font=normalsize,labelfont=sf,textfont=sf]{subfig}
\else
  \usepackage[caption=false,font=footnotesize]{subfig}
\fi

\begin{document}

	%
	%
	%

\title{A FEM enhanced Transfer Matrix method for optical grating design}
\author{Clara Zaccaria, Mattia Mancinelli, Lorenzo Pavesi~\IEEEmembership{Fellow,~IEEE}
		\thanks{C. Zaccaria, M.Mancinelli and L. Pavesi are with the Nanoscience Laboratory, Department
	of Physics, University of Trento, Povo (Trento), I-38123 Italy e-mail:clara.zaccaria@unitn.it.}
\thanks{Manuscript received \today; revised}}

\markboth{Journal of \LaTeX\ Class Files,}%
{Zaccaria \MakeLowercase{\textit{et al.}}: Bare Demo of IEEEtran.cls for IEEE Journals}

\maketitle

\begin{abstract}
A method to design gratings in integrated photonics, is presented. The method is based on a transfer matrix formalism enhanced by Finite Element Method (FEM) parameter calculations. The main advantages of the proposed technique are the easy of use, the fast optimization time and the versatility of the approach. Few examples of optimized gratings to obtain various scattered light field profiles for different applications are presented: a double-Gaussian profile, a flat top square profile, a spot profile on a chip surface, profiles suited to get efficient and selective coupling to single mode and multimode fibers. A discussion of the limits of the method and some insights on how to improve it are also discussed.
\end{abstract}


\IEEEpeerreviewmaketitle

\section{Introduction}
In integrated optics, gratings are widely used for coupling light in and out of a chip or to direct light beam in selected directions, such as in phased arrays \cite{Quaranta:resonant, McLamb:extr, Ziel:planar}. Optimization of the grating efficiency is done by controlling the geometrical parameters of the gratings \cite{Silberstain:int, khalid:apo, ML:capo,Chen:coup, Patri:coup, Nambiar:coup, Tang:coup, Dirk:coup, bridging}. Initially, periodic grating structures were used, then double etched and apodized grating structures were proposed and, finally, aperiodic gratings have been developed with the aim to increase the grating adequacy to the desired function. Nowadays, optimization techniques and machine learning approaches are widely used to design gratings \cite{ML:capo,ML:mapping}. Optimization techniques are based on simulators based on the physical model of the system and on optimization algorithms (particle swarm optimizer, genetic algorithm, gradient-based methods, ...). They have the advantage of a solid theoretical basis and a proper physical description of the whole system. However, they are typically computational intensive, requiring tens of minutes to complete a single grating simulation \cite{tempi:coup}.
On the other hand, machine learning approaches optimize structures with feed-forward or reservoir-computing neuronal networks after a training phase based on data-sets previously computed by physics-based simulators \cite{ML:capo, ML:mapping}. These methods are very effective (ms computational time) but are like black-boxes, i.e. physical insights on the optimum configuration are not easy. In addition, the machine learning solvers can have reliable performances mostly within the parameter range used during the training, which limits their application to very specific system. \\

In this paper, we propose a method to optimize grating geometries based on a transfer matrix (TM) formalism enhanced by a physical simulator. In this way we can get the physical insights typical of simulator based methods conserving the speed and the manageability of machine learning based approaches.\\

The paper is organized as follows. Section 2 presents the method based on TM coupled to a FEM (finite element method) simulator. Section 3 discusses the improvement in computational time. Section 4 reports many different examples of the proposed method: the creation of specific profiles at any distance from the grating, the realization of specific patterns on the surface of photonic chips for biological applications and the design of efficient fiber couplers for both single mode and multimode fibers. Section 5 discusses the limits and possible improvements of the method. Section 6 concludes the paper.

\section{The method to determine the field scattered by a grating} \label{TranferMatrix}

The goal of the method is to find the optimum layer sequence in a grating to obtain a target intensity profile of the diffracted beam. Therefore, we have to compute the spatial map of the scattered light intensity and, specifically, the profile of the scattered intensity at a given distance $y$ from a grating. Note that the method, due to the reciprocity of the optical path, can be also used to model a grating to input the light in a chip. We called our method \textit{FEM enhanced Tansfer Matrix} because we start from a FEM simulation of the basic elements of the system (named \textit{blocks} in the following) in order to get the reflected, transmitted and scattered fields by each block; these quantities are then used in the TM simulation in order to compute the behaviour of the whole grating.

\begin{figure}[!t]
\centering
\includegraphics[width=8.7cm]{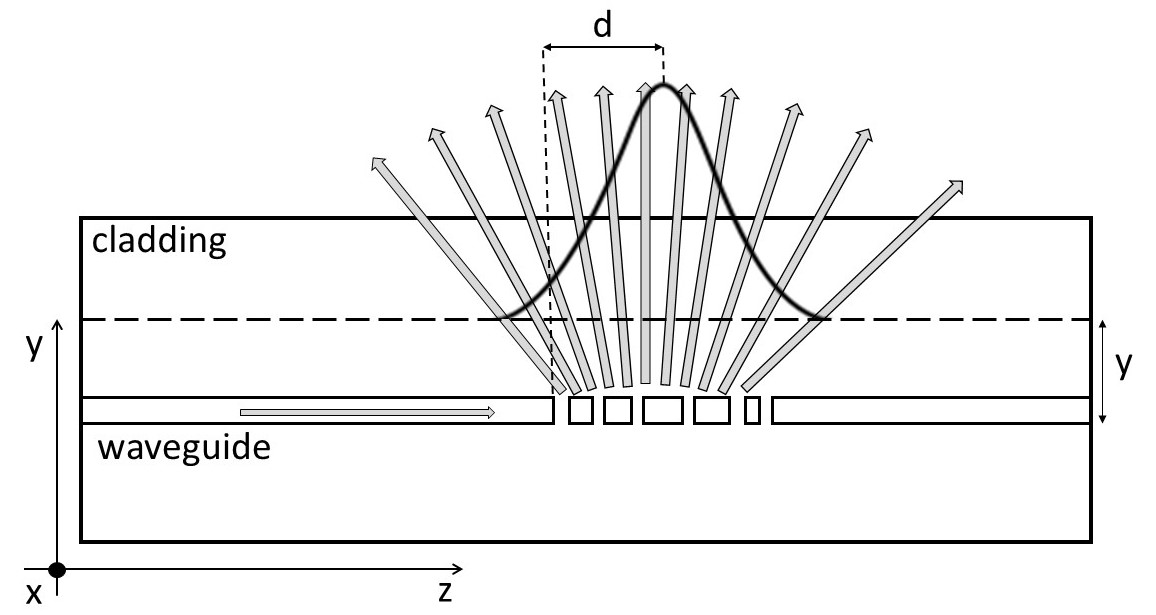}
\caption{Sketch of a deep etched grating which shows the two dimensional simulation system we are considering. Light propagating in a single mode waveguide embedded in a cladding layer, is scattered (arrows) by a grating whose layer lengths are different. At a certain distance $y$ from the grating in the y direction the scattered light should have a certain distribution, whose profile is here drawn by the thick line as a Gaussian profile. The center of the Gaussian is displaced from the grating by a distance d.}
\label{scatt}
\end{figure}

\begin{figure}[!t]
\centering
\includegraphics[width=8.7cm]{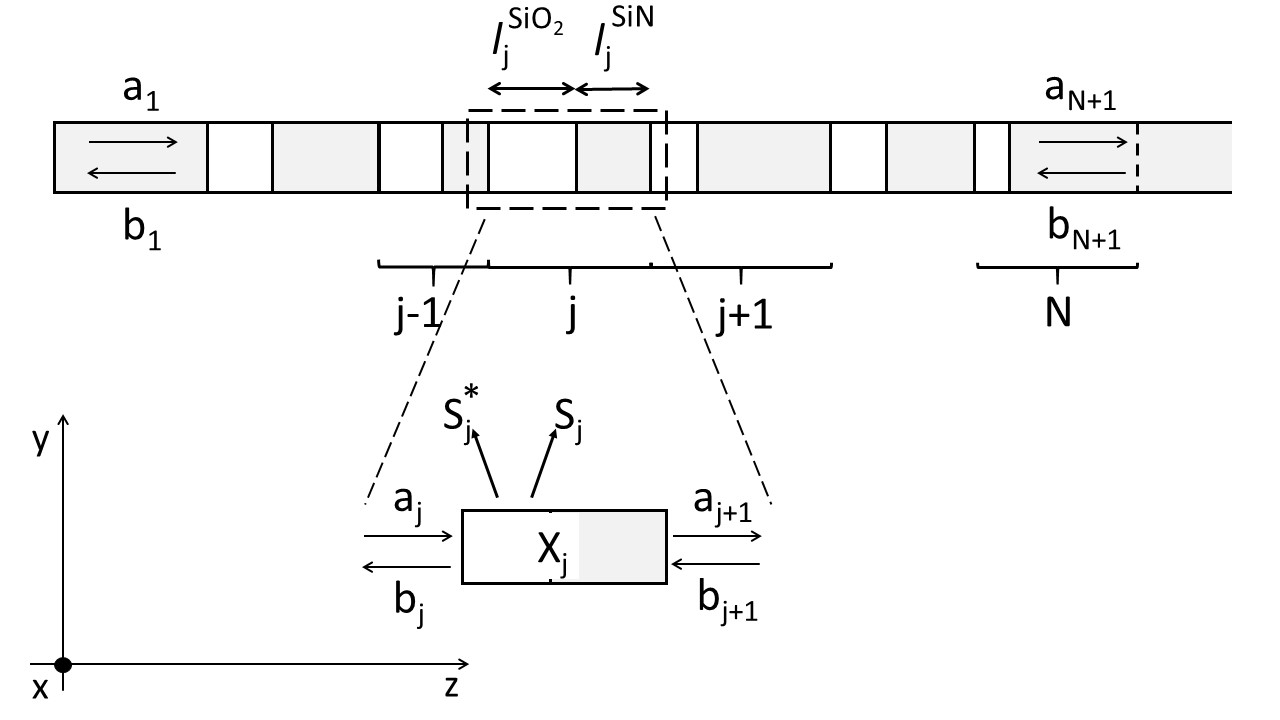}
\caption{Sketch of the layer sequence in the grating and definitions of the grating input waveguide modes (a$_1$, b$_1$), block indices ($j$), propagating (a$_j$) and back-propagating (b$_j$) waveguide modes, scattering fields ($S_j$ and $S^*_j$), single block transfer matrix ($X_j$), waveguide layer in the grating ($l_j^{SiN}$), scattering layer in the grating ($l_j^{SiO_2}$), and grating transmitted (a$_{N+1}$) waveguide mode. A SiN waveguide embedded in SiO$_2$ is considered.}
\label{sys}
\end{figure}

Let us consider a two dimensional approximation, where a grating is formed along a planar waveguide (Fig. \ref{scatt}). To model the system we use the TM method since it allows to easily compute the optical field propagation along the grating. In this discussion, we assume a SiN waveguide embedded into a SiO$_2$ cladding layer. Specifically, as shown in Fig. \ref{sys}, we divide the grating in blocks formed by a SiO$_2$ layer ($l^{SiO_2}_j$ long) and a SiN layer ($l^{SiN}_j$ long). Each block might have a different $l^{SiO_2}_j$ and $l^{SiN}_j$. Each block can transmit, reflect or scatter the incident light (see inset in Fig. \ref{sys}). The idea is to model the spatial profile of the light scattered by the grating  as a coherent superposition (interference) of the fields scattered by each individual block (Fig. \ref{scatt}).

The proposed model is based on the following assumptions:\\
1) each single block acts as a scatterer and the scattering depends only on the SiO$_2$ length;\\
2) the individual blocks are independent;\\
3) the scattering profile of the single block is calculated only once with a FEM simulator for any different SiO$_2$ layer length;\\
4) the reflection and transmission coefficients at the block interfaces are calculated by the same FEM simulation run used for the scattering profile calculation;\\
5) the propagating field is a waveguide propagation optical mode;\\
6) propagation losses in the grating are neglected (i.e. no losses in the SiN layers, no absorption in the SiO$_2$ layer).\\

Let us assume that the waveguide optical mode propagation occurs along the $z$-axis. Then, the mode in the grating region might have two components, the forward propagating optical mode $a_j$ and the backward propagating optical mode $b_j$. The propagation through the $j$-th block is described by a 2x2 matrix $X_j$ that links the modes (a$_j$,b$_j$) at the input with the modes (a$_{j+1}$,b$_{j+1}$) at the output of the $j$-th block. 

\begin{equation}
	\bigl(
	\begin{matrix}
		a_{j+1} \\
		b_{j+1}
	\end{matrix}
	\bigr)=X_j \bigl(
	\begin{matrix}
		a_j \\
		b_j
	\end{matrix}
	\bigr).
\end{equation}

The matrix $X_j$ can be obtained by the product of two matrices ($M_j$ and $F_j$) that describe the action of the SiO$_2$ layer  (matrix $M_j$) on the mode (i.e. reflection, scattering and transmission) and the phase accumulated by the mode while propagating in the SiN layer (matrix $F_j$). In details:

\begin{equation}
	X_j= F_j M_j,
\end{equation} 

\begin{equation}
M_j=\frac{1}{t^j_{21}}\begin{pmatrix}
t^j_{12}t^j_{21}-r^j_{12}r^j_{21} & r^j_{21} \\
-r^j_{12} & 1

\end{pmatrix} ,
\label{M}
\end{equation}

\begin{equation}
F=\begin{pmatrix}
	e^{-i\phi_j} & 0 \\
	0 & e^{i\phi_j} 
\end{pmatrix},
\label{F}
\end{equation}
where $\phi=n_{eff}k_0 l^{SiN}_j$.  
In \eqref{M}, $t^j_{12}$ and $r^j_{12}$ are respectively the transmission and reflection of the SiO$_2$ layer for a forward propagating mode, while $t^j_{21}$ and $r^j_{21}$ are the same coefficients for a backward propagating mode. Since optical reciprocity, $t^j_{12}=t^j_{21}=t^j$ and $r^j_{12}=r^j_{21}=r^j$. Note that these coefficients are different from the usual Fresnel coefficients, which depend only on the interface. Indeed, they consider the influence of the scattered field on the reflected and transmitted modes. Therefore, they are different for each $l_j^{SiO_2}$.
In \eqref{F}, $n_{eff}$ is the effective refractive index of the input waveguide mode $a_1$, $k_0 = 2 \pi /\lambda$ and $\lambda$ the wavelength.\\ 

Considering a grating formed by N different periods (blocks in our model), we can reconstruct the field distribution inside the grating by using the TM multiplication

\begin{equation}
	\bigl(
	\begin{matrix}
		a_{N+1} \\
		b_{N+1}
	\end{matrix}
	\bigr)=X_{N} \bigl(
	\begin{matrix}
		a_{N} \\
		b_{N}
	\end{matrix}
	\bigr)= \ldots = \prod_{j=1}^{N} X_{j} \bigl(
	\begin{matrix}
		a_{1} \\
		b_{1}
	\end{matrix}
	\bigr),
\end{equation}
where $a_1=1$ is the normalized optical mode entering the grating, and $a_{N+1}$ is the optical mode transmitted by the grating. We impose $b_{N+1}=0$ since the grating is terminated by a waveguide where the mode propagates only in the forward direction (see Fig. \ref{sys}).

Knowing the field components inside the grating, the overall scattered intensity by the grating ($I_g(y,z)$) can be computed as the overlap (interference) of the fields scattered by each single block (E$_j(y,z)$), i.e.
\begin{equation}
	I_g(y,z)= \left| \sum_{j=1}^{N} E_j(y,z)\right|^2, \label{sommatoria}
\end{equation}

\noindent where

\begin{equation}
	E_j(y,z)=a_j S_j(y,z) + b_{j+1}S_j^*(y,z)e^{i\phi_j}.
	\label{E}
\end{equation}

Note that $E_j(y,z)$ and $\sum_{j=1}^{N} E_j(y,z)$ are complex quantities: therefore, they give access to the amplitude and the phase of the scattered fields.

In \eqref{E}, $S_j(y,z)$ and $S_j^*(y,z)$ are the scattering fields from the $j-th$ block, for a forward and a backward propagating optical mode, respectively (see Fig. \ref{sys}). Again,  $S_j(y,z)$ and $S_j^*(y,z)$ depend on $l_j^{SiO_2}$.

\begin{figure}[!t]
    \centering
    \includegraphics[width=8.5cm]{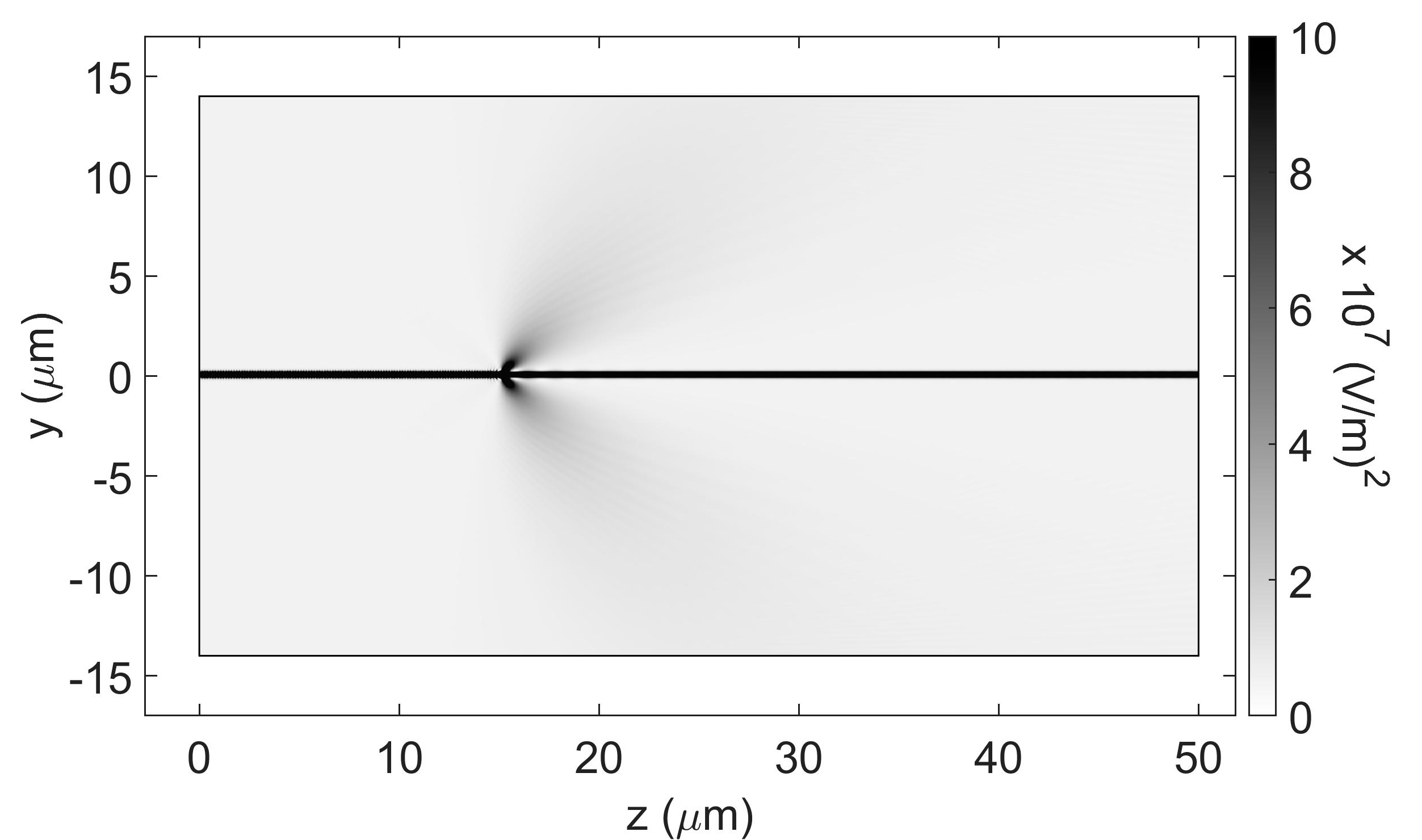}
    \includegraphics[width=8.5cm]{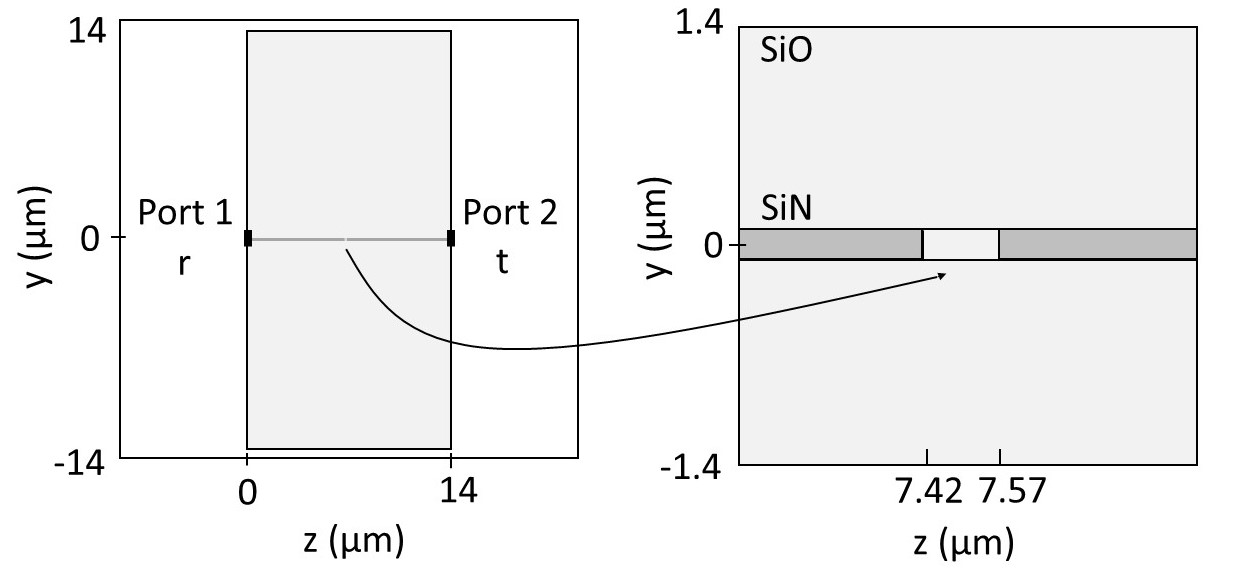}
    \caption{(top) FEM simulation domain to compute the single block parameters where a single SiO$_2$ layer is interposed in a SiN waveguide cladded by SiO$_2$. Here the SiN waveguide is 150 nm high and the SiO$_2$ layer is 250 nm long. The two thick vertical lines (not in scale) show the positions where the fields are computed to extract the $r^j$ (Port 1) and $t^j$ (Port 2) parameters. (bottom) Scattering map for a forward propagating optical mode. The dark horizontal line refers to the optical mode which propagates in the waveguide. The gray scale refers to the field intensity and is given on the right of the panel. The simulation is done for a 488 nm wavelength and with the parameters shown in the top panel, a part a longer waveguide.}
    \label{Comsol1}
\end{figure}

In order to get the values for the $r^j$ and $t^j$ coefficients and the maps of the scattering fields ($S_j(y,z)$), we performed FEM simulations of a single SiO$_2$ layer of length $l_j^{SiO_2}$ interposed in a SiN waveguide (Fig. \ref{Comsol1}). We did our simulations in the transverse electric (TE) polarization. 

We used scattering boundary conditions \cite{scattering_boundaries} except for the input and output ports, and we calculated $r^j$ and $t^j$ as
\begin{equation}
	r^j=\frac{\text{optical mode reflected to port 1}}{\text{input optical mode}},
\end{equation}
\begin{equation}
     t^j=\frac{\text{optical mode transmitted to port 2}}{\text{input optical mode}}.
\end{equation}

We corrected their phases with the phase shift given by the SiN length. With the same simulation, $S_j(y,z)$ is computed (Fig. \ref{Comsol1} bottom). Then, the $z$-coordinate is properly shifted to the actual $z$ position of the $j$-th block. $S^*_j(y,z)$ is simply obtained by reflecting $z \longrightarrow -z$ to represent a backward propagating field scattered by the SiO$_2$ layer.

The FEM simulations are repeated for the various SiO$_2$ layer thicknesses and a database of the various coefficients for various block parameters is built. To save computational time, we split the FEM simulation in two runs: in the first we compute $r^j$ and $t^j$ with a small simulation domain and a high resolution mesh, while in the second we used a large simulation domain with a low mesh resolution to compute $S_j(y,z)$. 

\begin{figure*}[t!]
	\centering
	\subfloat[]{\includegraphics[scale=0.072]{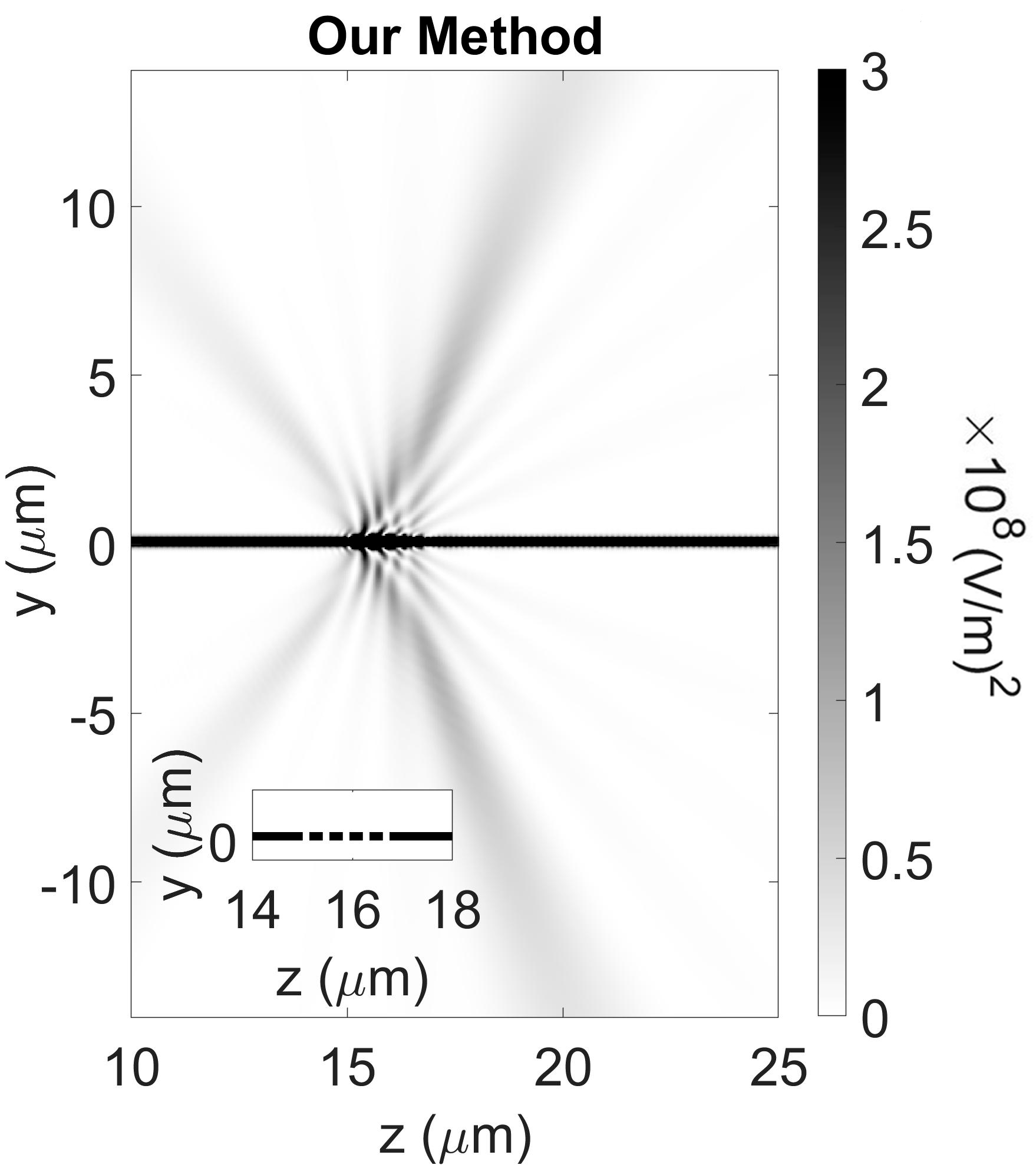}}
		\hfil
	\subfloat[]{\includegraphics[scale=0.073]{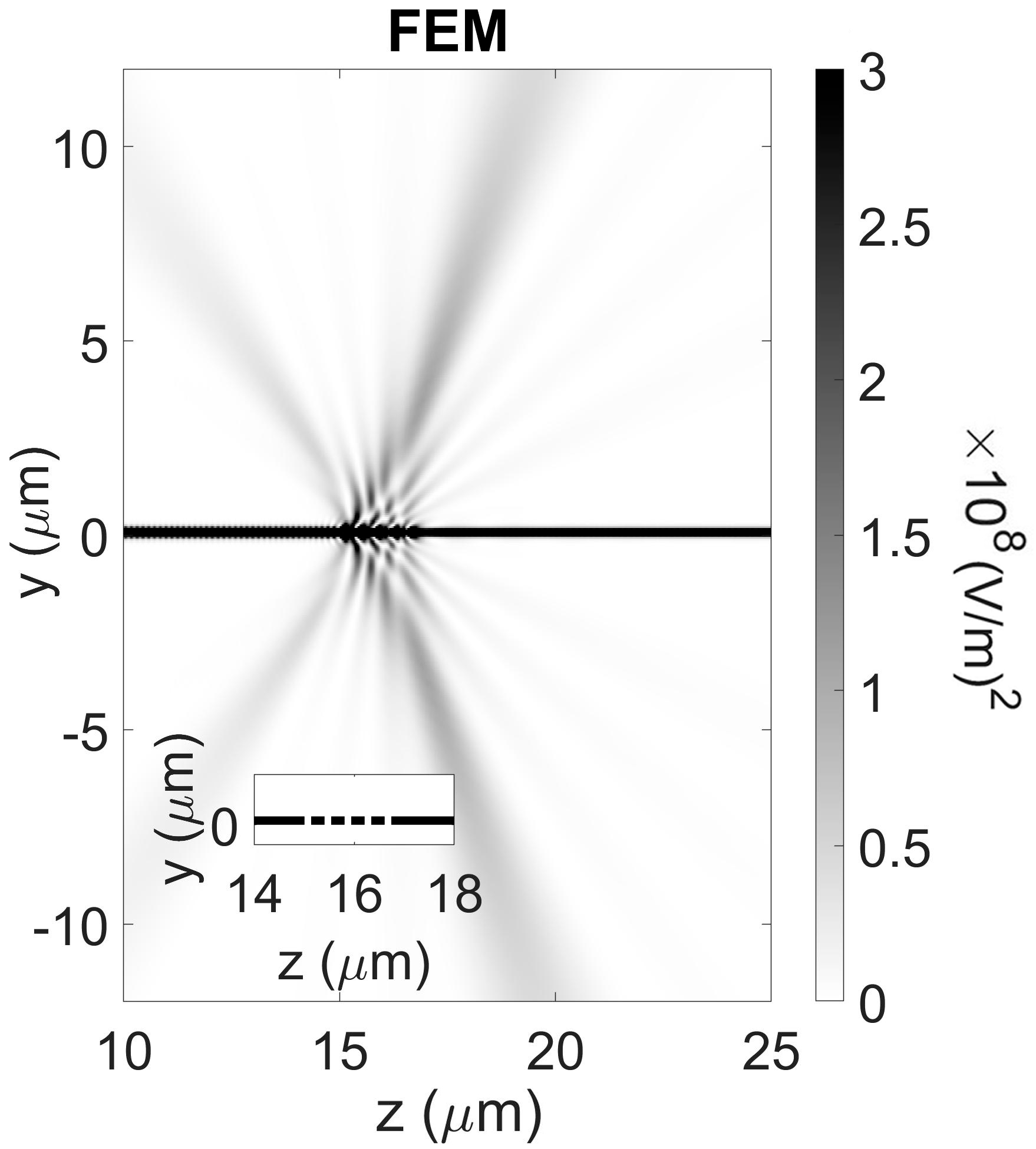}}
		\hfil
	\subfloat[]{\includegraphics[scale=0.072]{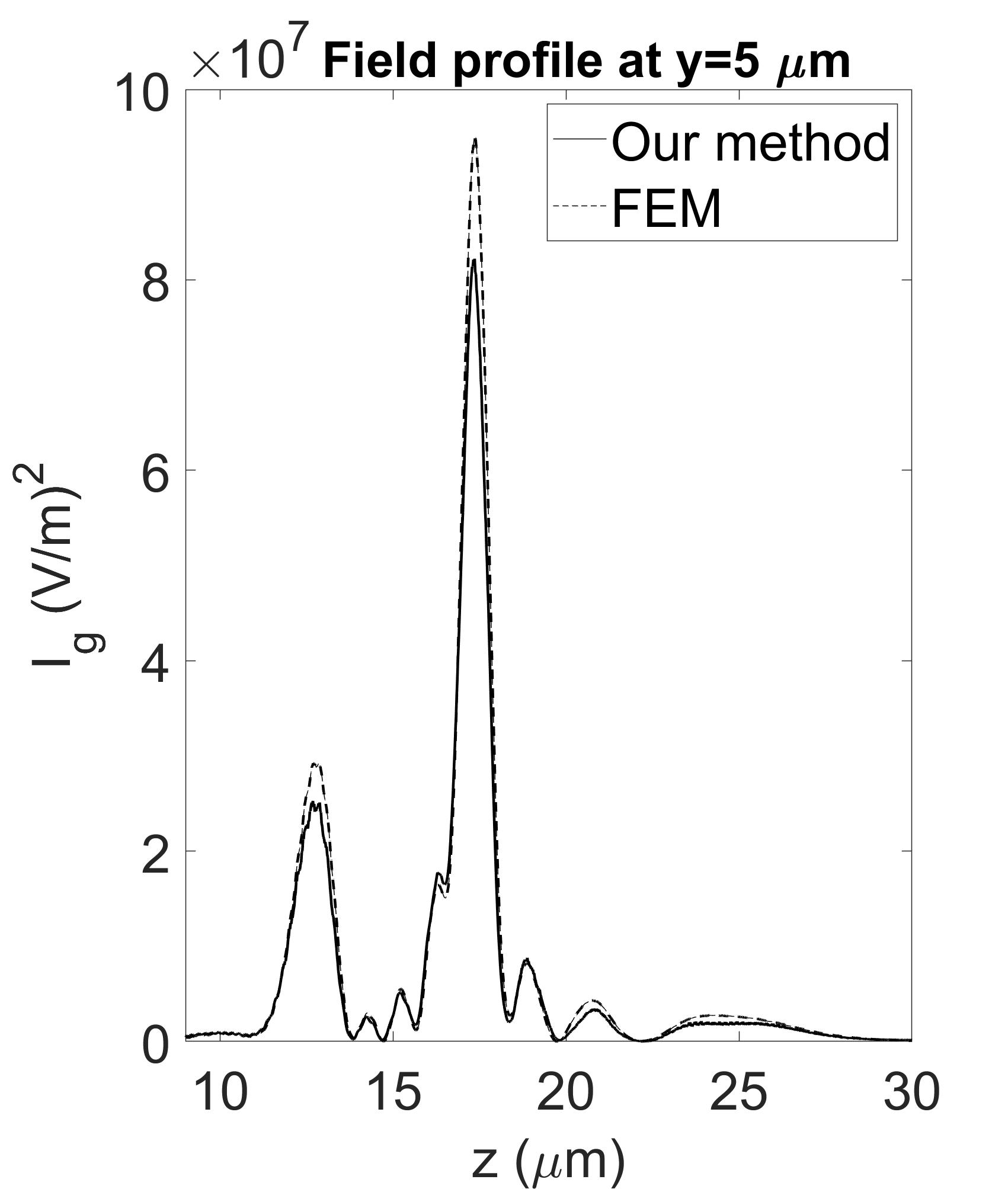}}
		\hfil
	\caption{Comparison between our method (a) and a FEM simulation (b) of $I_g(y,z)$ from a periodic grating made by N=5 blocks of 250 nm long SiN layer and 150 nm long SiO$_2$ layer. The SiN waveguide is 150nm high and the light wavelength is 488 nm. The dark horizontal line refers to the optical mode which propagates in the waveguide. The gray scale refers to the field intensity and is given on the right of the panel. The insets show a sketch of the periodic N=5 grating. (c) The $I_g(y,z)$ profiles obtained with our method (full line) and the FEM simulation (dashed line) at y=5 $\mu$m. }
	\label{confronto}
\end{figure*}

The comparison of the TM based simulation results with those computed with a full 2D FEM grating simulation is shown in Fig. \ref{confronto}. This allows checking the accuracy of the proposed method. It is observed that the $I_g(y,z)$ profiles are equal, only a small difference due to the use of the scattering boundary conditions in the FEM simulation is observed. Indeed, in the scattering boundary conditions only normal incidence fields are transmitted at the boundaries, for all other incidence angles the fields are partially reflected and contribute to $I_g(y,z)$.

\begin{table*}[h!] 
	\centering
	\begin{tabular}{l|cc|cc|cc}
		\toprule
	  	   & \multicolumn{2}{c|}{FEM} &\multicolumn{4}{c}{our method} \\
	  	   & \multicolumn{2}{c|}{map} &\multicolumn{2}{c|}{map} &\multicolumn{2}{c}{profile}\\
	     N & 1 iteration & optimization& 1 iteration & optimization& 1 iteration & optimization  \\
		\midrule
		1 & 26 s & -  & -       & -       & -         & - \\
		5 & 26 s & 4.2 days & 1.22 s & 4.8 h & 0.017 s& 4 min \\
		10 & 26 s& 5.8 days & 2.66 s & 14.2 h & 0.020 s & 6.4 min \\
		20 & 26 s& 19.4 days & 5.64 s & 4.2 days & 0.025 s & 27 min \\
		\bottomrule
	\end{tabular}
	\caption{Comparison between the duration of a periodic grating simulation with a FEM simulator and our method (for a full scattering map or a simple scattering profile). For each method, the time needed for one iteration and for the optimization are shown. The optimization time corresponds to the time that the optimization algorithm, runned with our method, with 2N+1 parameters, takes to perform 14100 iterations for N=5, 19300 iterations for N=10 and 64700 iterations for N=20, respectively. The same number of iterations are used for the FEM calculations for the sake of comparison. The computations were performed on a i5-1035G4 CPU with 8 GB RAM. The time values of the single interations are the averages performed on 4 simulations to take into account the time spent by the CPU on the underlying system processes: the error in the time values is of the order of few ms.}
	\label{tempi}
\end{table*}

\section{About the computational time}

One of the main advantage of our method is its computational efficiency which is instrumental in the optimization of the grating layer lengths. The benchmarks are the typical duration times of the more widely used methods: FEM and FDTD (finite difference time domain) methods require tens of minutes to complete a single grating simulation\cite{tempi:coup, tempi:coup1}, while a fully vectorial eigenmodes expansion and propagation tool (CAMFR) reaches tens of seconds \cite{tempi:coup, tempi:coup1, tempi:camfr1, tempi:camfr2}. These times transform to days or weeks when the optimization of the grating has to explore a wide parameter space. Our method requires FEM calculations to build the parameters table and, then, a simple and fast TM calculation to optimize the grating structure. This results in a significantly faster optimization than the fully FEM optimization (see Tab. \ref{tempi}). In our experiments, a grating optimization requires only few hours on a laptop with a i5-1035G4 CPU with 8 GB RAM.

More quantitatively, Table \ref{tempi} shows the computation times that the aforementioned laptop takes to simulate a N=1, N=5, N=10, N=20 blocks periodic grating with a FEM (Comsol Multiphysics 5.3a) and with our method (implemented in Matlab 2019b). The simulations performed for this comparison use the same spatial domain of the N=1 FEM simulation. Since the domain and the mesh used in the FEM are fixed, the computational times do not change increasing N for the FEM for one iteration. Instead, for our method, the computational time increases by increasing N. On the other hand, when the number of iterations increases (as in an optimization run) our method significantly outperforms the FEM calculations. This is observed even considering the initial time needed to compute the grating layer parameters. A further advantage of our method is the possibility to shorten the computation time by simply computing the scattering profile at a given distance from the grating.

\section{Optimization of the grating geometric parameters to produce any scattering maps}

The FEM enhanced TM method we just described can be used to compute the optimum grating parameters that produce the target $I^T_g(y,z)$. The specific optimization algorithm depends on the problem one addresses. In the examples presented in the following, we used the particle swarm optimization algorithm (PSO) because it is freely available in most commercial computation suites, it is fast, it is able to handle many optimization parameters and it is widely used in structure optimizations \cite{part_swarm, part_swarm1, part_swarm2, ML:mapping, ML:capo}.
Known the target function $I^T_g(y,z)$, the optimization algorithm has to minimize the cost function $\Xi$. This last depends on the specific problem. To speed up the simulations, we included the reflection from the first block and the transmission from the last one in $\Xi$ only when they are larger than 1$\%$.

\subsection{Example 1: a scattered profile}

As a first example, we optimize a grating to create a desired distribution of scattered light at a distance from the grating plane. Here the cost function is defined as the standard deviation of the obtained distribution $I_g(y,z)$ from the desired one $I^T_g(y,z)$ plus the transmitted $T=|a_{N+1}|^2$ and reflected $R=|b_1|^2$ light from the grating, i.e.
\begin{equation}
	\Xi=\sqrt{\int_{space} (I^T_g(y,z)- I_g(y,z))^2 dydz }+R +T. \label{cost}
\end{equation}

Since we are interested in the shape of our scattered distribution, both $I_g(y,z)$ and $I^T_g(y,z)$ are normalized to 1, while the two last terms are introduced in order to minimize the waveguide transmission and reflection, i.e to maximize the scattered light from the grating. The surface integral in \eqref{cost} is replaced by a line integral if only the scattering profile at a given height is of concern.

The grating is based on a 220 nm high Si waveguide (refractive index n=3.478) working at 1550 nm, embedded in SiO$_2$ (n=1.446). The parameters to be optimized are the layer lengths ($l_{SiO_2}$ and $l_{Si}$) in the grating, the distance of the profile from the grating (d, computed from the position of the first grating layer) and the height $y$ from the grating plane at which the desired profile is formed (Fig. \ref{scatt}). We varied $l_{SiO_2}$ in the range 130-200 nm, while $l_{Si}$ between 130 nm and $\lambda/n_{eff}$. This last choice restricts the phase change in the Si layers to 2$\pi$. If longer Si thicknesses are used, the optimization algorithm is no longer able to differentiate between thicknesses with 2$\pi$ phase differences. We left also the distances $d$ and $y$ free to move in a range 0-30 $\mu$m and 2-35 $\mu$m. The first block starts at z=50 $\mu$m. 

\begin{figure}
	\centering
	\subfloat[\label{bigaN5}]{\includegraphics[scale=0.072]{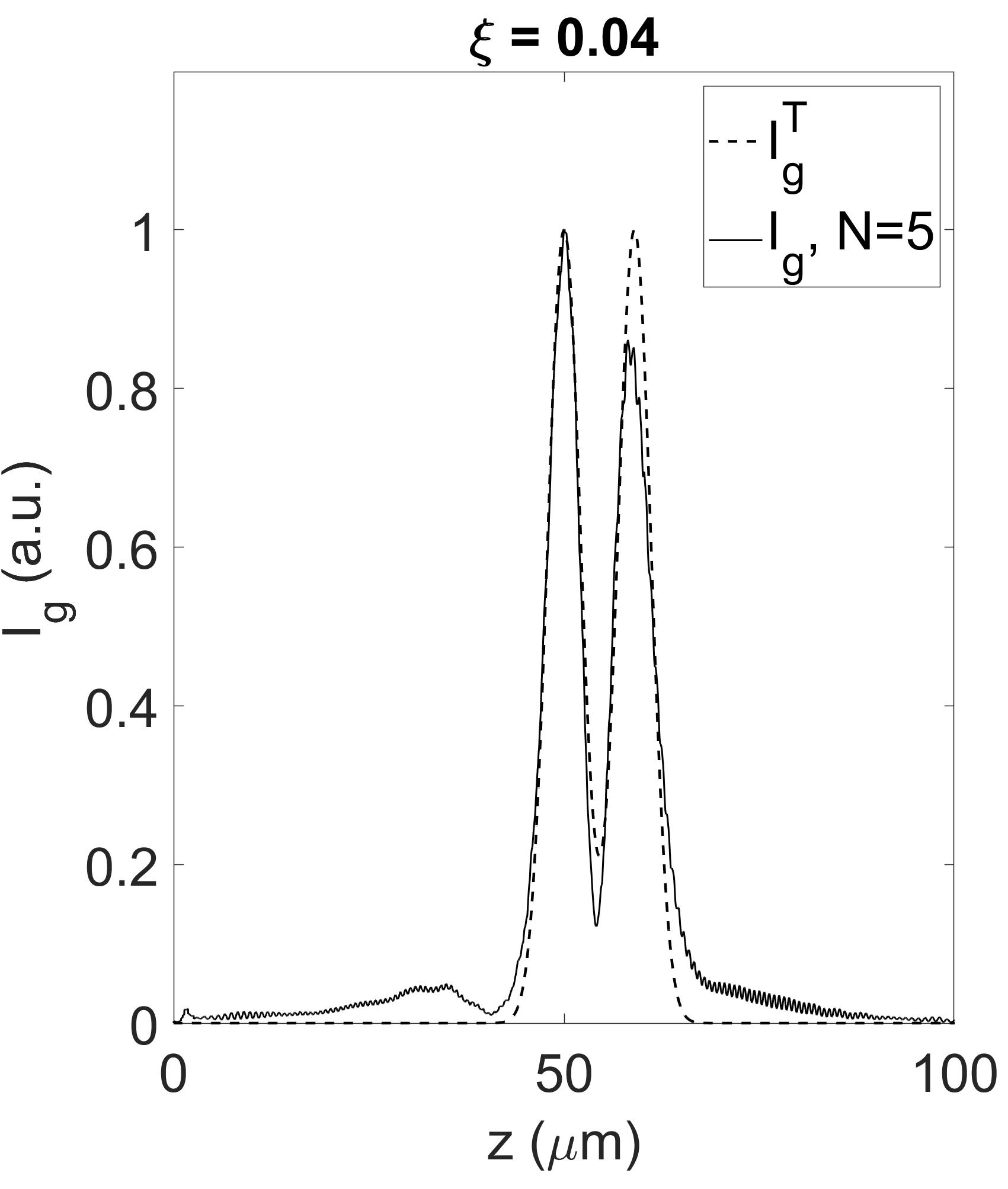}}
	\subfloat[\label{bigaN20}]{\includegraphics[scale=0.072]{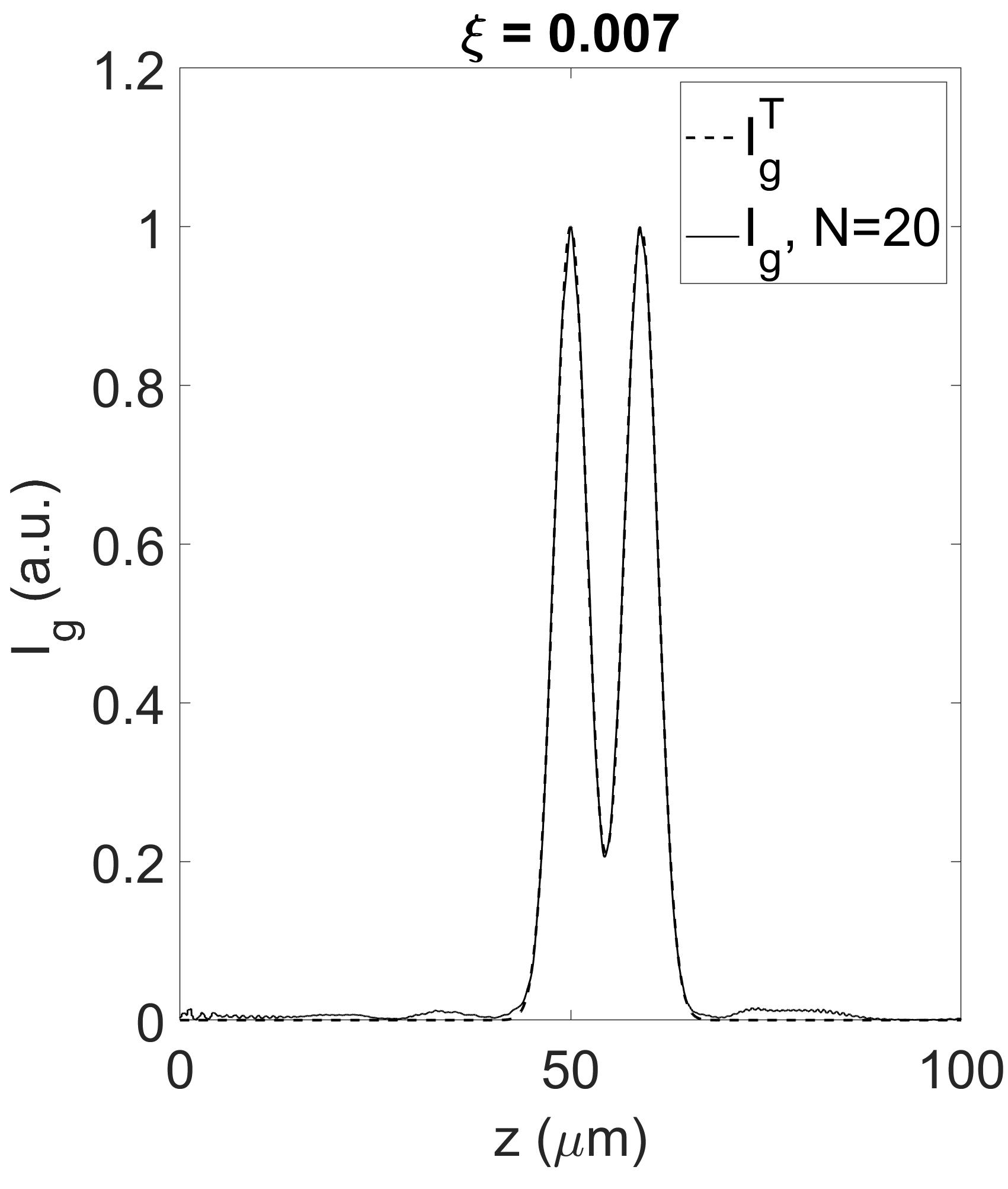}}\\
	\subfloat[\label{bigaN40}]{\includegraphics[scale=0.072]{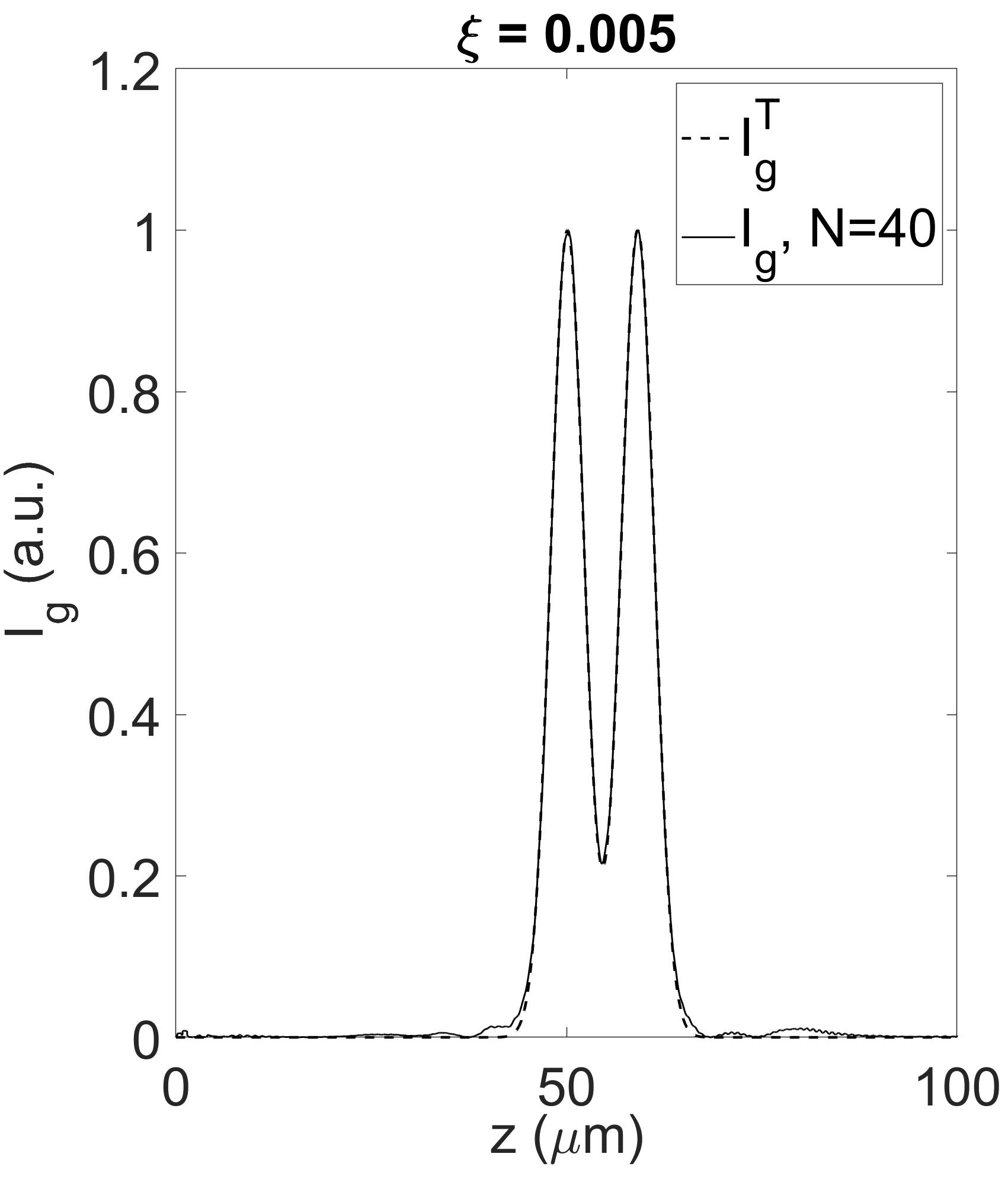}}
	\subfloat[\label{N20_2D}]{\includegraphics[scale=0.072]{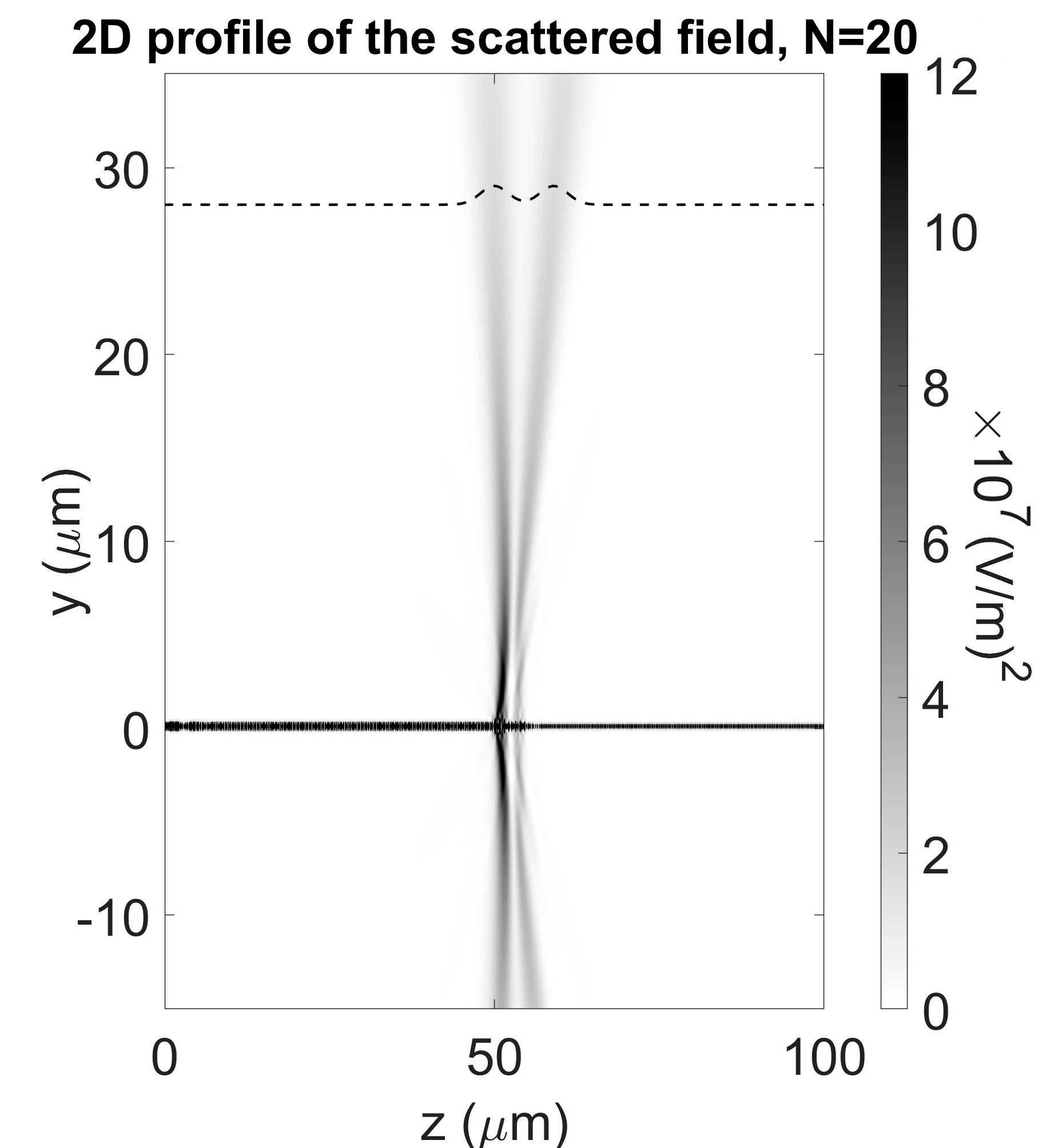}}\\
	\subfloat[\label{N_D}]{\includegraphics[width=8 cm]{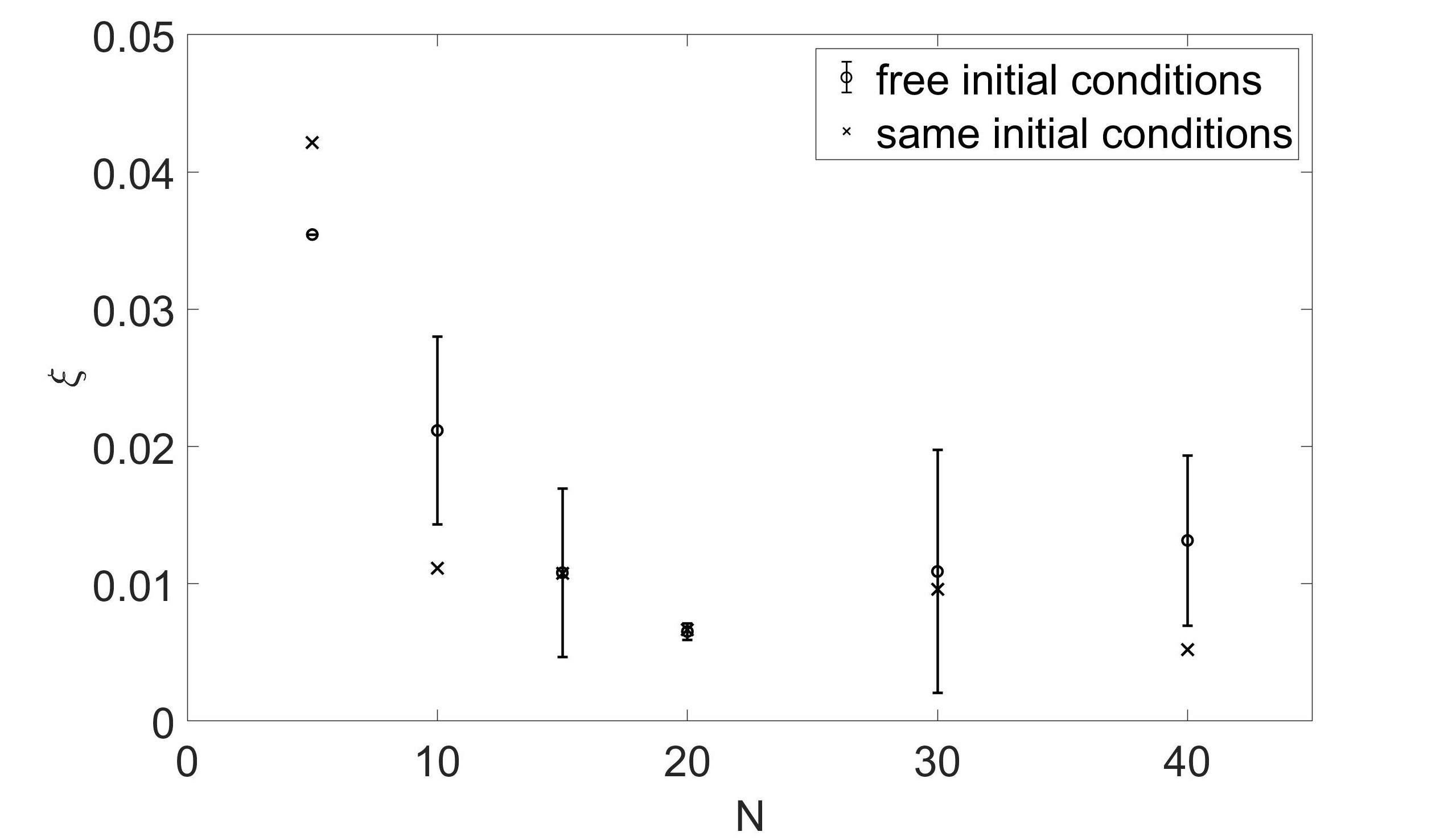}}	\\
	\subfloat[\label{grat_N20}]{\includegraphics[width=8 cm]{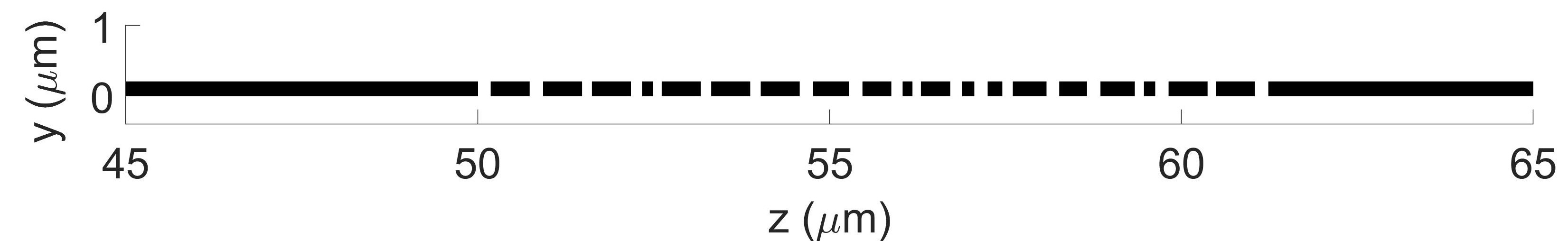}}
	\caption{Grating to produce a double-Gaussian profile. These results are obtained by the proposed method. (a), (b) and (c): The scattered profiles (full line) compared to the target profile (dashed line) for N=5, N=20 and N=40 blocks, respectively. (d) Scattering map for N=20. The dotted line shows the profile at the best height $y$. The dark horizontal line refers to the optical mode which propagates in the waveguide. The gray scale refers to the field intensity and is given on the right of the panel. (e) The $\xi$ value as a function of N. Circles for a free starting condition, crosses for a fixed starting condition. Error bars are due to multiple optimization runs.  (f) The layer sequence for the optimized N=20 grating.}
	\label{biga}
\end{figure}

In a first case, the target profile is a double-Gaussian profile (two gaussians with 5 $\mu$m FWHM (full-width at half maximum) and 9 $\mu$m of separation between the maxima).
Figure \ref{bigaN5} shows the optimized grating with N=5. The double-Gaussian profile is obtained  at y=15.9 $\mu$m and d=0 $\mu$m. Comparing this with the target profile, we note differences on the peak heights and the presence of unwanted side peaks. We decided then to perform a quantitative study on the influence of increasing N on the scattered profile. We define $\xi=\sqrt{\int (I^T_g(z)- I_g(z))^2 dz }/\int I^T_g(z) dz$ as an index that estimates, in percentage, the difference between the obtained and the target profiles. Figure \ref{N_D} shows $\xi$ as a function of N. We let the optimization initial conditions free (circles in Fig. \ref{N_D}), and we got large errors bars (calculated over 4 different optimization runs). In fact, the PSO algorithm finds many different minima for $\Xi$ which limits the stability and the reproducibility of the solutions. Error bars are significantly reduced if the optimization starts from the same initial conditions (crosses in Fig. \ref{N_D}). Increasing N, $\xi$ reduces (i.e. better double-Gaussian profiles, see Fig. \ref{bigaN20}, \ref{bigaN40}). In addition, increasing N, $R$ and $T$ become smaller: this increases the scattering efficiency of the grating. Interestingly, $\xi$ saturates as a function of N at N=20 (Fig. \ref{N_D}). This is due to the fact that most of the light is scattered by the first 20 blocks and only a weak signal is propagating in the grating after the 20-th block. In fact, a simple estimate shows that less than 2\% of the light is transmitted after the 20-th block \footnote{the shorter 130 nm long SiO$_2$ layer has a transmission coefficient of 0.82; after 20 equal periods with N=20, the transmission reduces to $(0.82)^{20}=0.02$}. Increasing N, the PSO available phase-space widens, which results in an over-fitting situation. In this case, $\xi$ no longer decreases (Fig. \ref{N_D}).\\

\begin{figure}[!t]	
	\centering
	\subfloat[\label{quadN40}]{\includegraphics[scale=0.07]{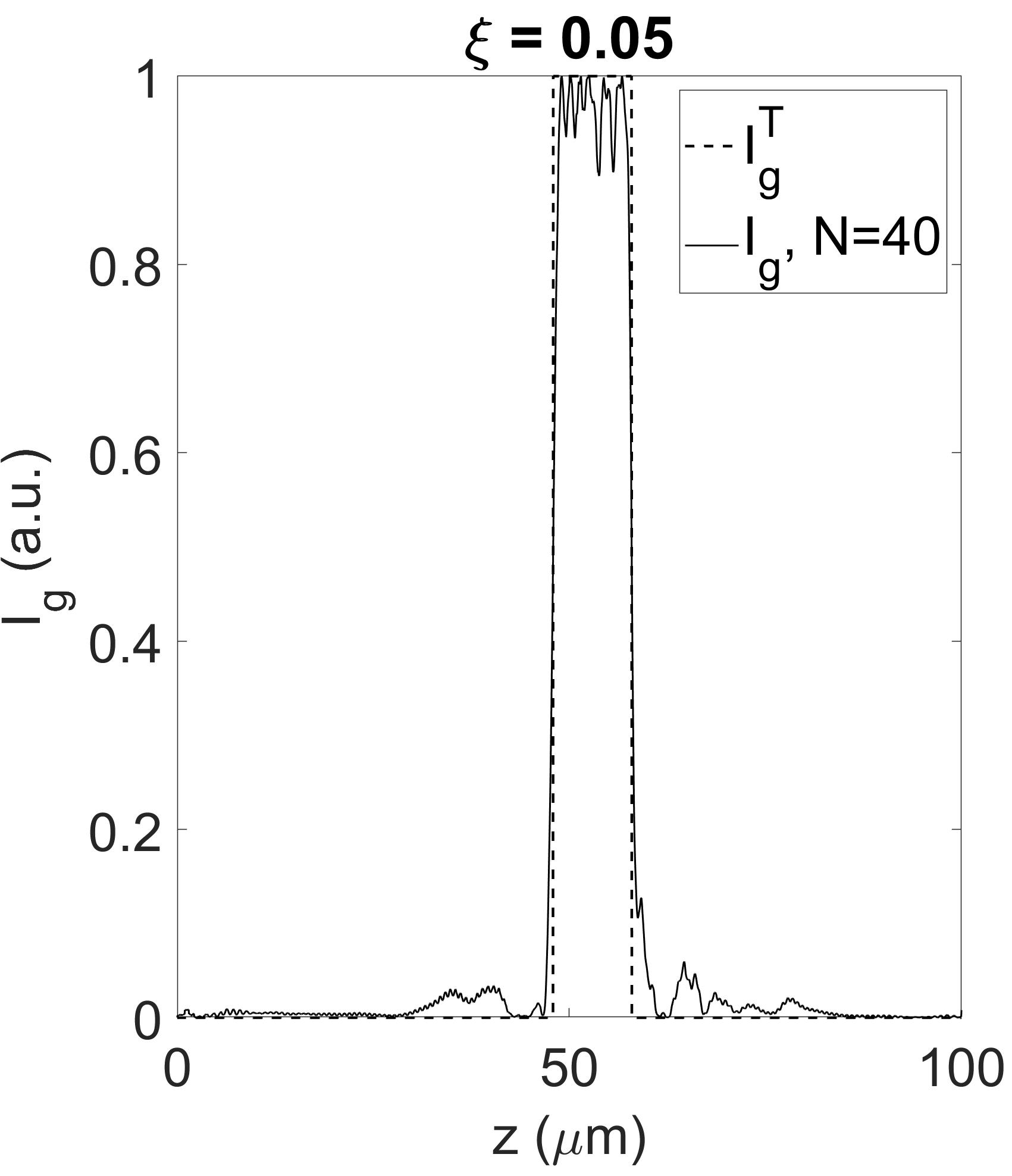}}
	\subfloat[\label{quadN40_2D}]{\includegraphics[scale=0.07]{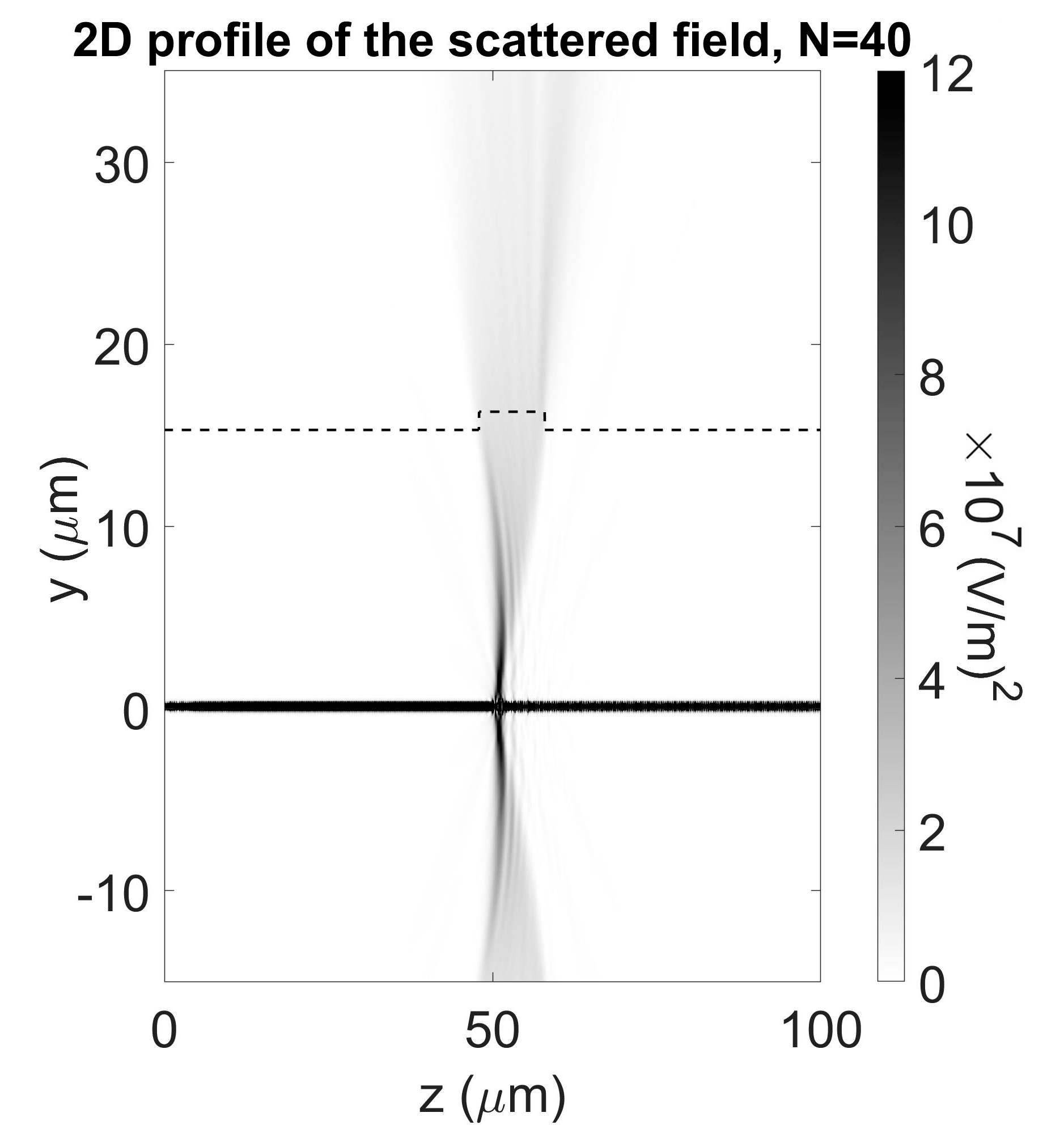}}\\
	\subfloat[\label{grat_N40}]{\includegraphics[width=8.5cm]{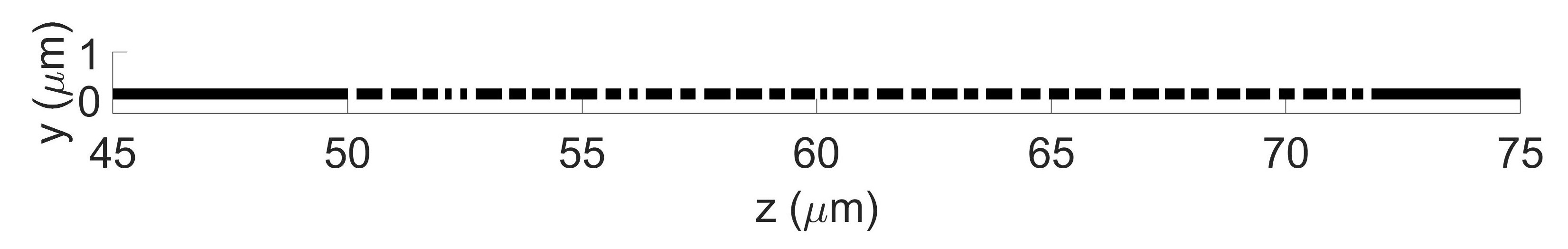}}	
	\caption{Optimized grating which yields a flat top square profile. (a) The scattered profile compared to the desired profile for N=40. (b) Scattering map for N=40. The dotted line shows the profile at the best height y. The dark horizontal line refers to the optical mode which propagates in the waveguide. The gray scale refers to the field intensity and is given on the right of the panel. (d) The layer sequence for the optimized N=40 grating. }
	\label{quadr}
\end{figure}

In a second case, the target profile is a 10 $\mu$m wide flat top square profile (Fig. \ref{quadr}). As in the previous case, $\xi$ decreases by increasing N without evidences of saturation. In this case, the optimum value of N has to be found as a trade-off between the computational time and the $\xi$ value. Note that the observed fluctuations on the top of the profile are due to interference.

\subsection{Example 2: spot on the surface of a photonic chip} \label{sezione_neurone}

In many applications, it is useful to have a light spot on the surface of a photonic chip. In optogenetics, one aims at exciting light sensitive proteins in transgenic neurons cultured on a photonic chip \cite{welkenhuysen2016integrated}. Therefore, we optimized a grating to produce a blue spot of the dimension of a typical neuron soma (10 $\mu$m) at a distance which matches the position of the neurons on the chip surface. In this case, the waveguide was a 150 nm high SiN waveguide (n=1.968) embedded in SiO$_2$, the working wavelength was 488 nm and the spot has to be formed at a distance of about y=5 $\mu$m from the waveguide. The target $I^T_g(y,z)$ is a Gaussian, with a 8 $\mu$m FWHM. The various parameters were free to move in a range 150-400 nm for $l_{SiO_2}$, 150 nm and $\lambda/n_{eff}$ for $l_{SiN}$ and 0-20 $\mu$m for d; $y$ was fixed to 5 $\mu$m.

\begin{figure}[h]
	\centering
	\subfloat[\label{202D}]{\includegraphics[scale=0.07]{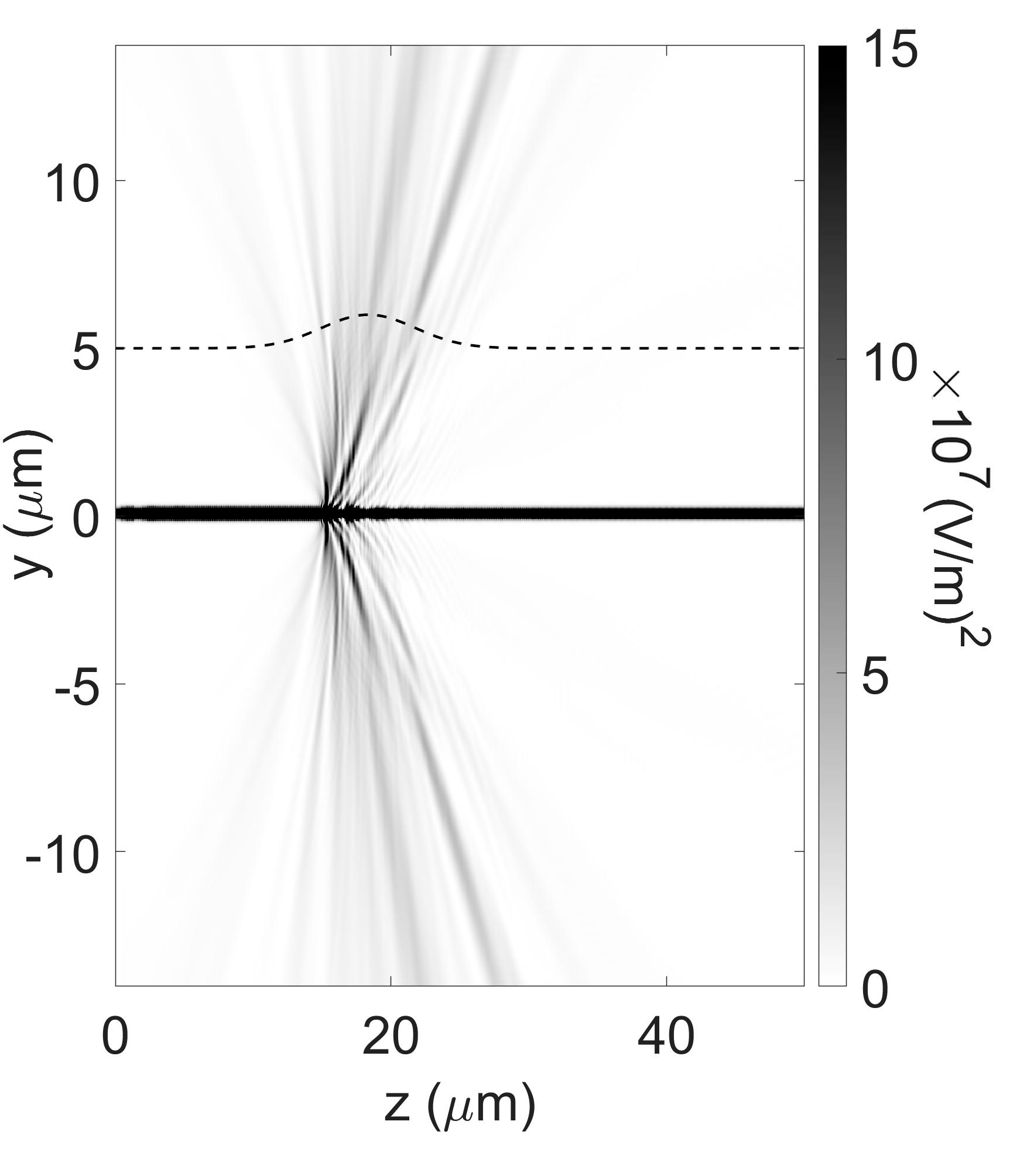}}
		\subfloat[\label{spot}]{\includegraphics[scale=0.07]{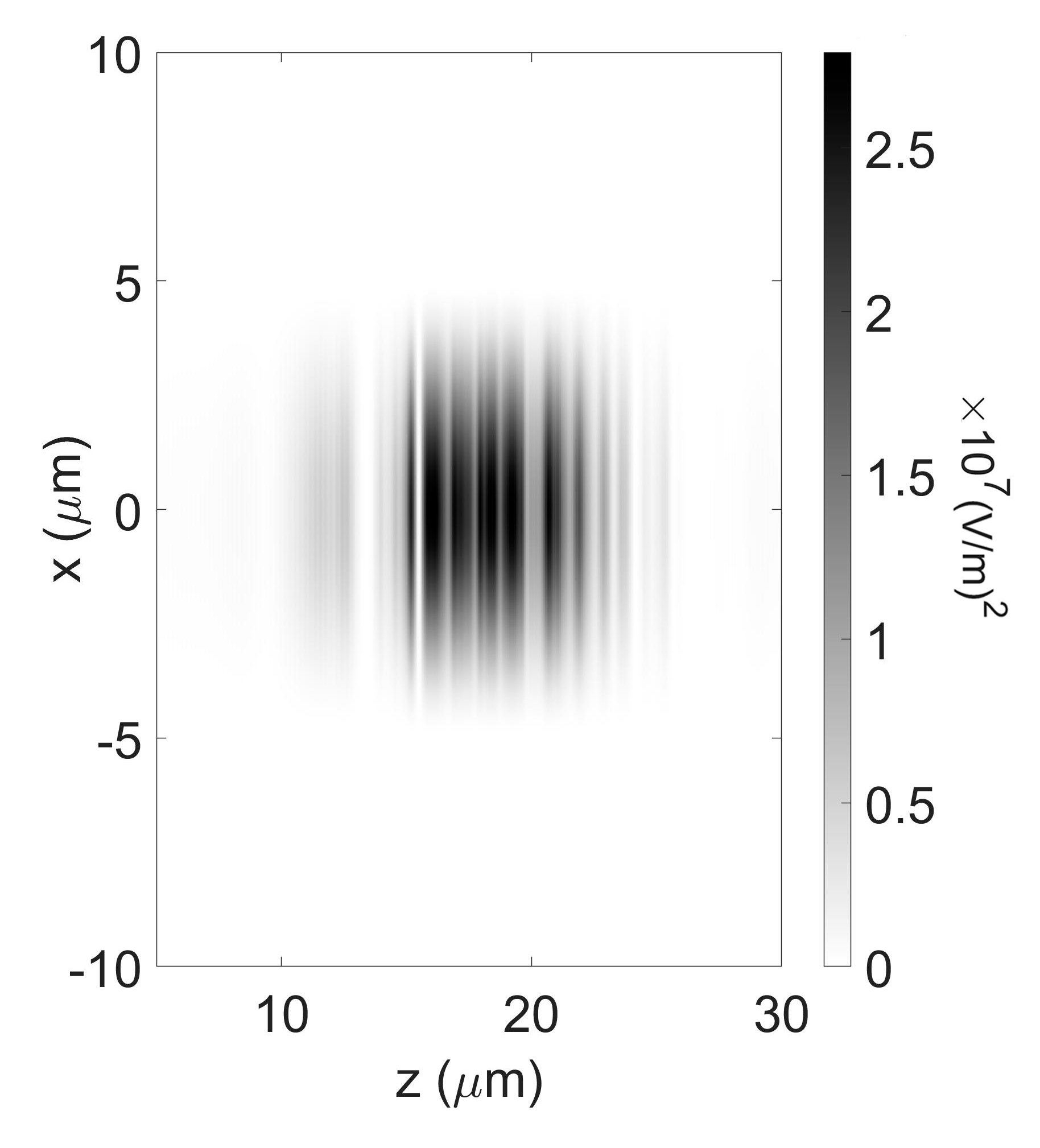}}\\
	\subfloat[\label{grating}]{\includegraphics[width=8.5 cm]{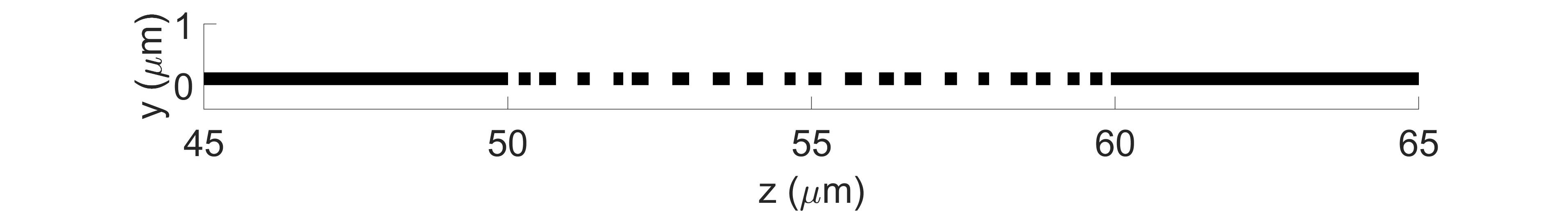}}
	\caption{(a) Scattering map for N=20. The dotted line shows the profile at y=5 $\mu$m. The dark horizontal line refer to the optical mode which propagates in the waveguide. The gray scale refers to the field intensity and is given on the right of the panel. (b) The spot on the surface of the photonic chip. The gray scale refers to the field intensity and is given on the right of the panel. (c) The layer sequence for the optimized N=20 grating.}
	\label{neur}
\end{figure}

The cost function for the PSO is given by \eqref{cost}. Once the grating is optimized, we define the grating efficiency $\eta$ as the percentage of the input power that is diffracted in the spot. If the integrated optical intensity on the spot is $I_{spot}$, then $\eta=\frac{1}{2} (1-R-T) \times I_{spot}/I_{0}$. Here, the $1/2$ term accounts for the symmetric up and down scattering (see Fig. \ref{Comsol1}), the term $(1-R-T)$ accounts for the input light which is reflected or transmitted by the grating, while the last term considers the fraction of scattered light $I_0$ which is actually concentrated in the spot. $I_0$ is calculated as the line integral of the scattered light at y=1 $\mu$m in order not to overlap with the evanescent field of the waveguide mode, while $I_{spot}$ is the integral of the scattered profile on a $z$ interval equal to the FWHM of the target distribution.
Based on $\xi$, which gets a minimum value between N=15 and N=20, we fixed N=20.
In Fig. \ref{202D} the obtained profile at y=5 $\mu$m and d=3.36 $\mu$m is shown. The obtained $\eta$ is 43\%. Only 5\% of the optical power is out of the spot. In Fig. \ref{spot},is plotted the spot that is created on the surface of the chip: this image is obtained by the multiplication of the fields propagating along $z$ and $x$ directions, independently obtained: the field along $z$ is the one calculated by our method, while the field distribution along $x$ results from a FEM based mode solver, considering a waveguide with the same materials and a rectangular cross section 8$\times$0.15 $\mu$m$^2$.

Note that the grating length is 9.935 $\mu$m: the grating optimization allows to concentrate most of the scattered intensity into a 10 $\mu$m wide spot.

\subsection{Example 3: grating couplers for monomodal and multimode fibers} \label{couplingpar}

\begin{figure}[h]
	\centering
	\includegraphics[width=8.5cm]{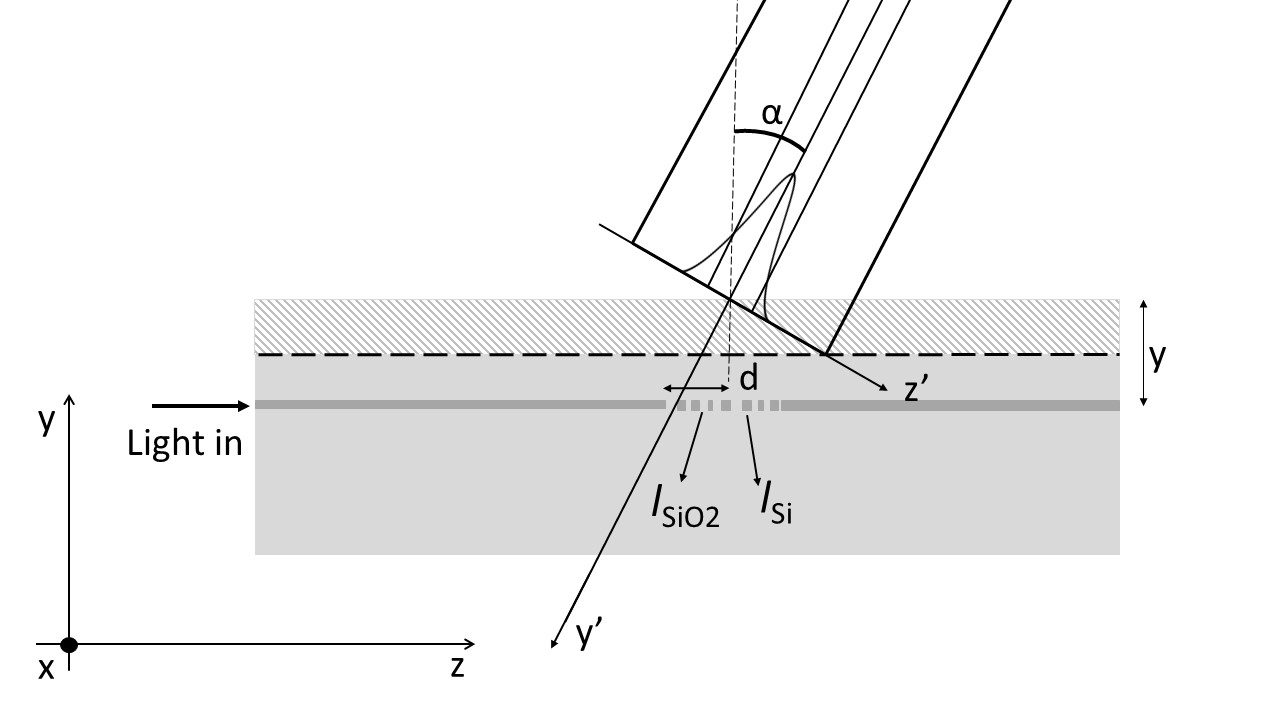}
	\caption{Sketch of the considered system.}
	\label{coup}
\end{figure}

One of the main use of gratings is to couple light in and out of a chip in and from a fiber. Therefore, the grating design aims at maximize the overlap between the scattered light and the modes of a fiber. The target profile might lay on a plane which can be tilted with respect to the waveguide plane. This provides a further optimization parameter, the angle $\alpha$ from the axis normal to the surface of the chip. Figure \ref{coup} defines the different parameters. In this example, we assume a Si waveguide (n=3.478) and a wavelength of 1550 nm, $l_{Si}$ varies between 130 nm and  $\lambda/n_{eff}$, $l_{SiO_2}$ between 130 and 200 nm, $d$ between 0.5 and 30 $\mu$m, $y$ between 2 and 35 $\mu$m (the entire simulation domain) and $\alpha$ between 0 and 45$^{\circ}$.\\

The cost function is defined as $\Xi=1-\eta_c$, where $\eta_c=1/2(1-R-T)\Gamma$ is the waveguide to fiber coupling efficiency. $\Gamma$ is the optical overlap integral between the complex scattered field ($\sum_{j=1}^{N} E_j(y,z)$) and the optical fiber mode \cite{Chang}. Since we ignore the $R$ and $T$ values as long as they are smaller than 1\%, we simplify $\Xi$ to: 
\begin{equation}
\Xi'= -\Gamma + R + T.
\label{cost_acc}
\end{equation}

\begin{figure}[!t]
	\centering
	\subfloat[]{\includegraphics[width=8cm]{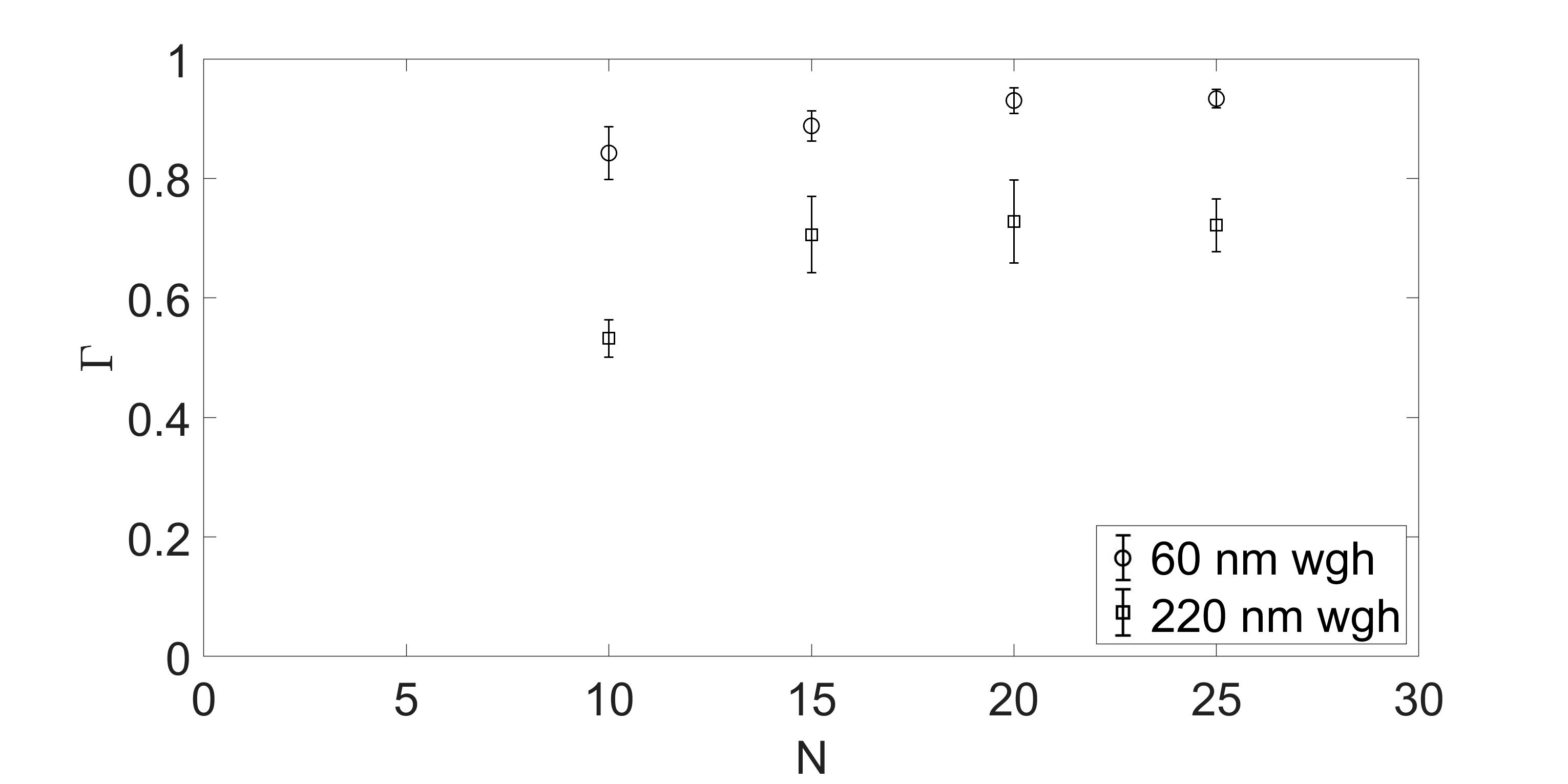}}\\
	\subfloat[]{\includegraphics[width=8cm]{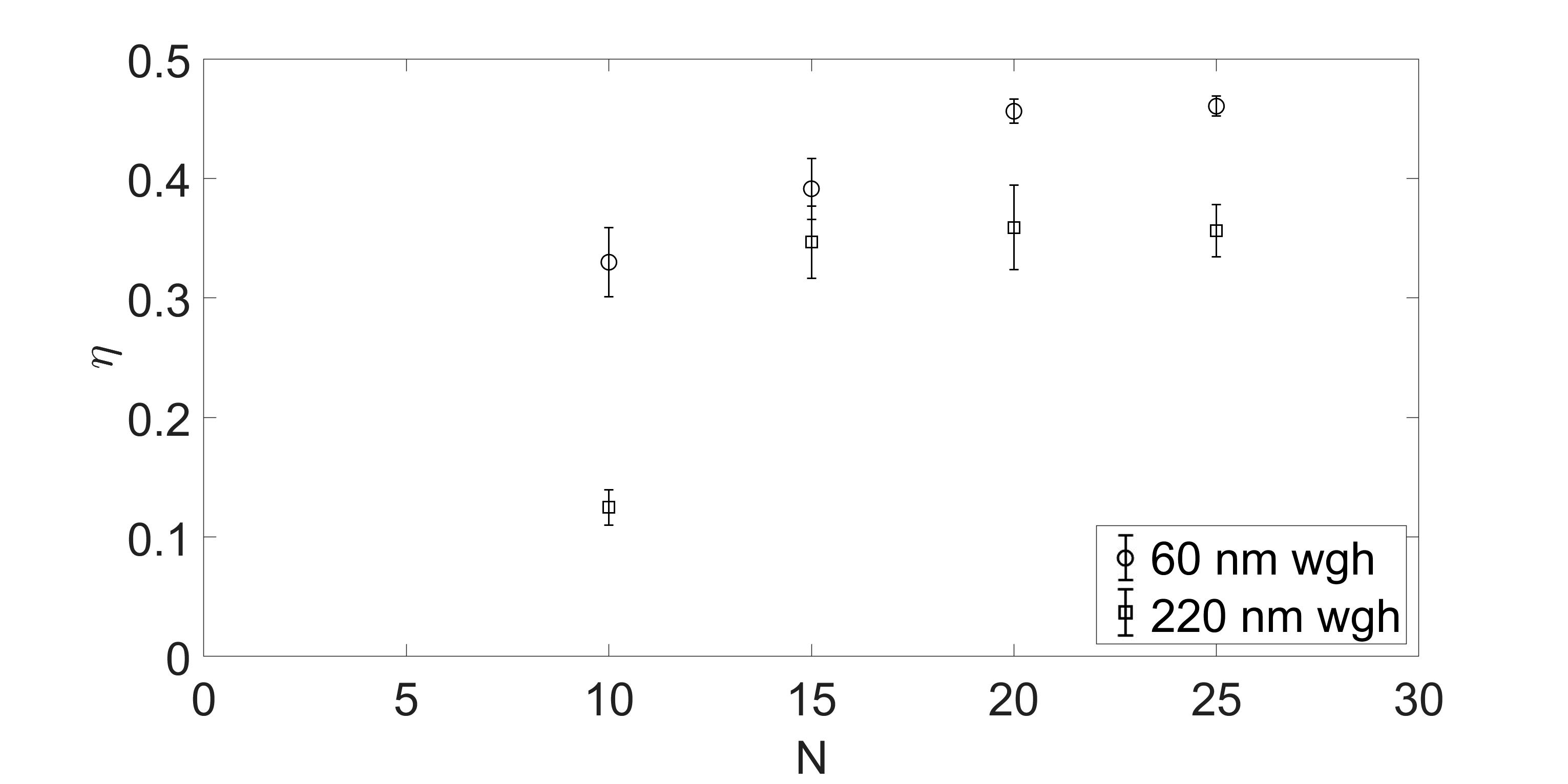}}
	\caption{Overlap integral (a) and coupling efficiency (b) between a Si waveguide and a single mode optical fiber as a function of the number N of blocks in the grating.}
	\label{high}
\end{figure}

Maximizing the overlap means minimizing the differences between the amplitudes and the phases of the scattered field and the optical fiber mode.
The first grating was optimized for a single mode fiber with a 9.5 $\mu$m core diameter. We considered two different Si waveguides of thicknesses equal to 220 nm and 60 nm, respectively. Figure \ref{high} shows $\Gamma$ and $\eta_c$ as a function of N for the two waveguides. Increasing N, both $\Gamma$ and $\eta_c$ increase. A larger efficiency is observed for the 60 nm thick waveguide than for the 200 nm thick one ($\Gamma= 95\%$ and $75\%$, respectively). Since $n_{eff}$ in the thin waveguide tends to the one of the silica cladding, the scattering from each block is reduced and more light propagates through the blocks. Therefore, all the N blocks are effective in the optimization phase, i.e. more parameters are available to the PSO to increase the overlap of the scattered field with the fiber mode.

\begin{figure}[!t]
	\centering
	\subfloat[\label{2Dmon}]{\includegraphics[width=9cm]{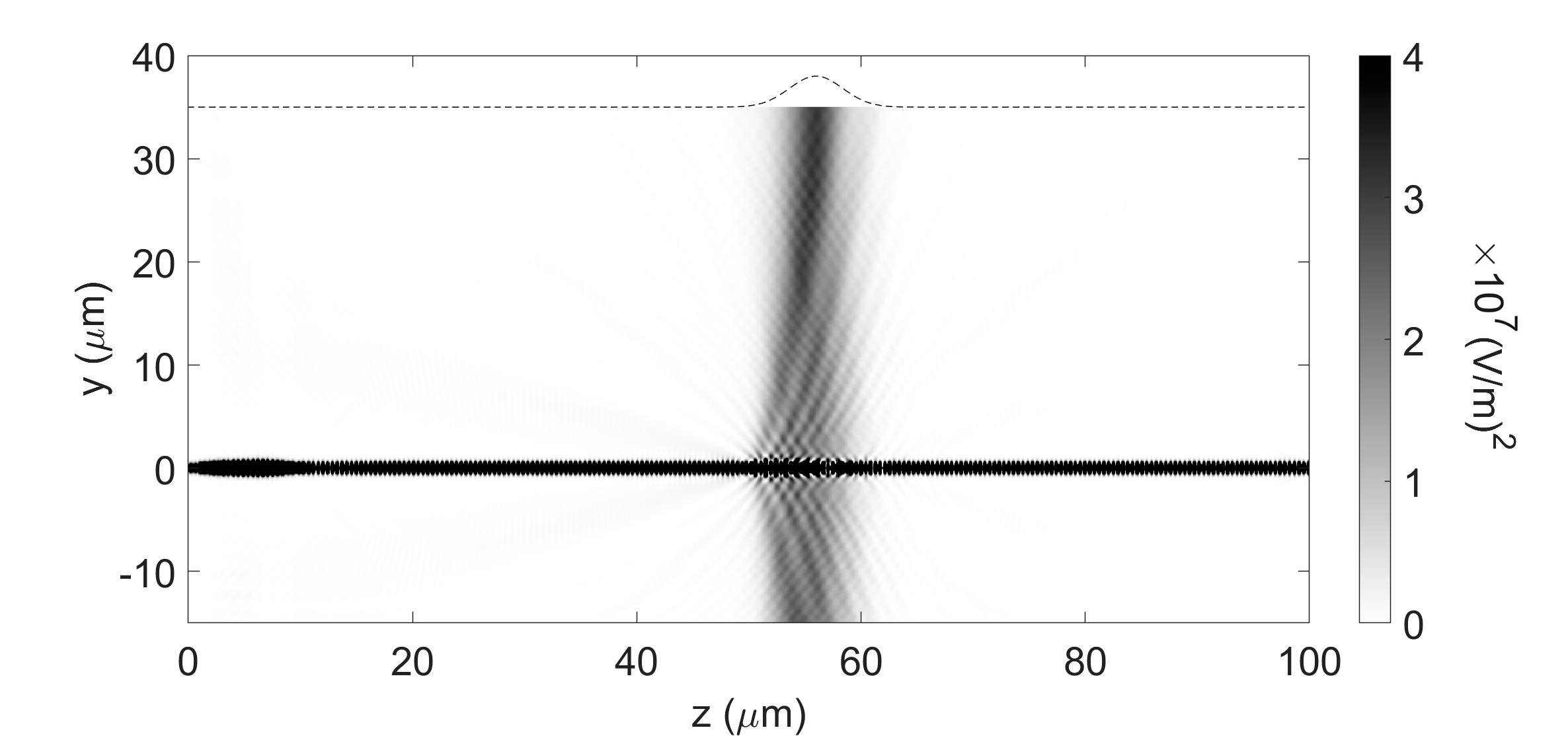}}\\
	\subfloat[\label{geom_mon}]{\includegraphics[width=9cm]{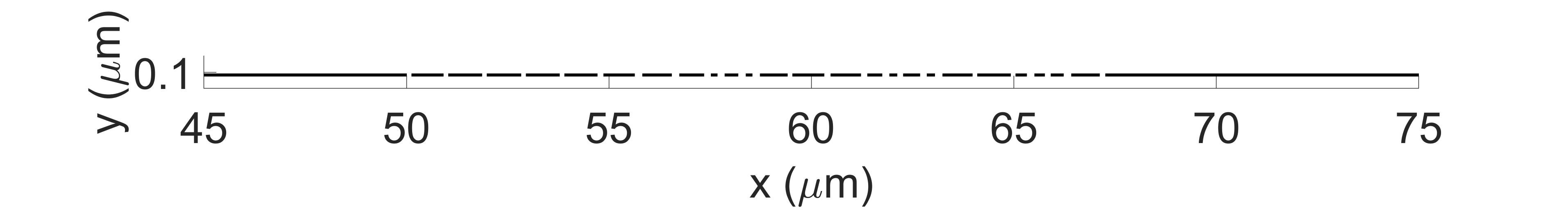}}
	\caption{(a) Scattering map from the optimized N=25 grating. (b) Layer sequence for the optimized aperiodic geometry.}
	\label{N25mon}
\end{figure}

A best coupling efficiency of 47\% is achieved after grating optimization (see Table \ref{valori}). Figure \ref{N25mon} shows the scattering map (i.e. $I_g$ distribution, defined as in formula \eqref{sommatoria}) and the actual sequence of layers. The dashed line represents the profile of the target field intensity $I^T$. The maximum efficiency is obtained for a fiber placed at a distance $y=$34.9 $\mu$m, centered at $d=$6 $\mu$m and tilted by an angle $\alpha$=1.46$^{\circ}$. In this case, a mode overlap of 95\% is obtained. Note that a periodic grating with 25 periods ($l_{SiO_2}$=160 nm and $l_{Si}$=760 nm) yields an efficiency of 7.95\%, with an overlap of 42.47\%, but an R equal to 0.6256.

The grating optimization can also be performed to design a grating which efficiently and selectively couples light to a specific mode of a multimode fiber. In this case we considered a thin Si waveguide (60 nm), and a multimode fiber with a 9 $\mu$m diameter core. We studied in particular the one-dimensional profiles of 4 Hermite Gaussian (HG) modes HG00, HG10, HG20 and HG30.
We run the optimization for each mode and the various geometries.

\begin{figure*}[h!]
	\centering
	\subfloat[]{\includegraphics[width=7.7cm]{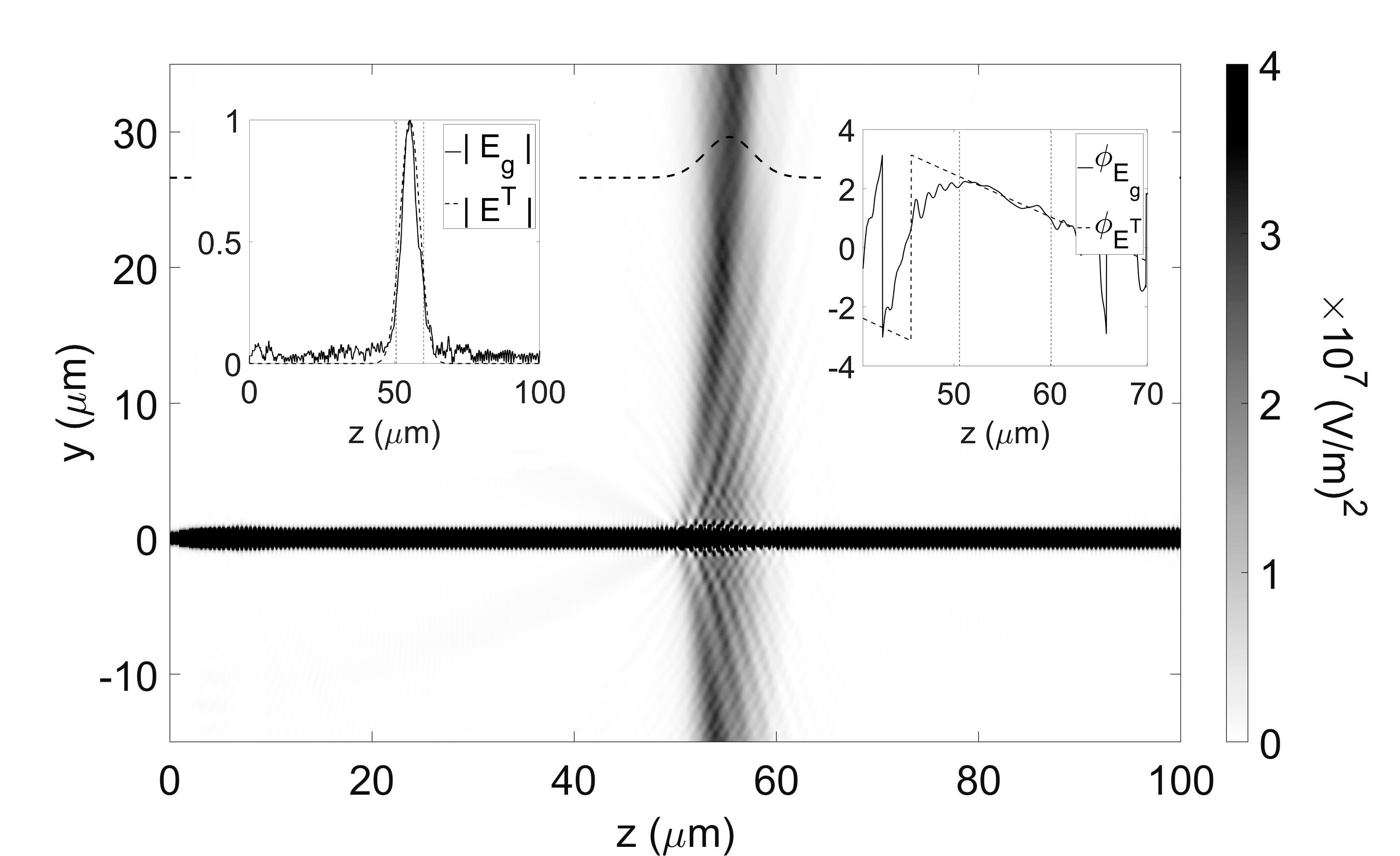}}
	\subfloat[]{\includegraphics[width=7.7cm]{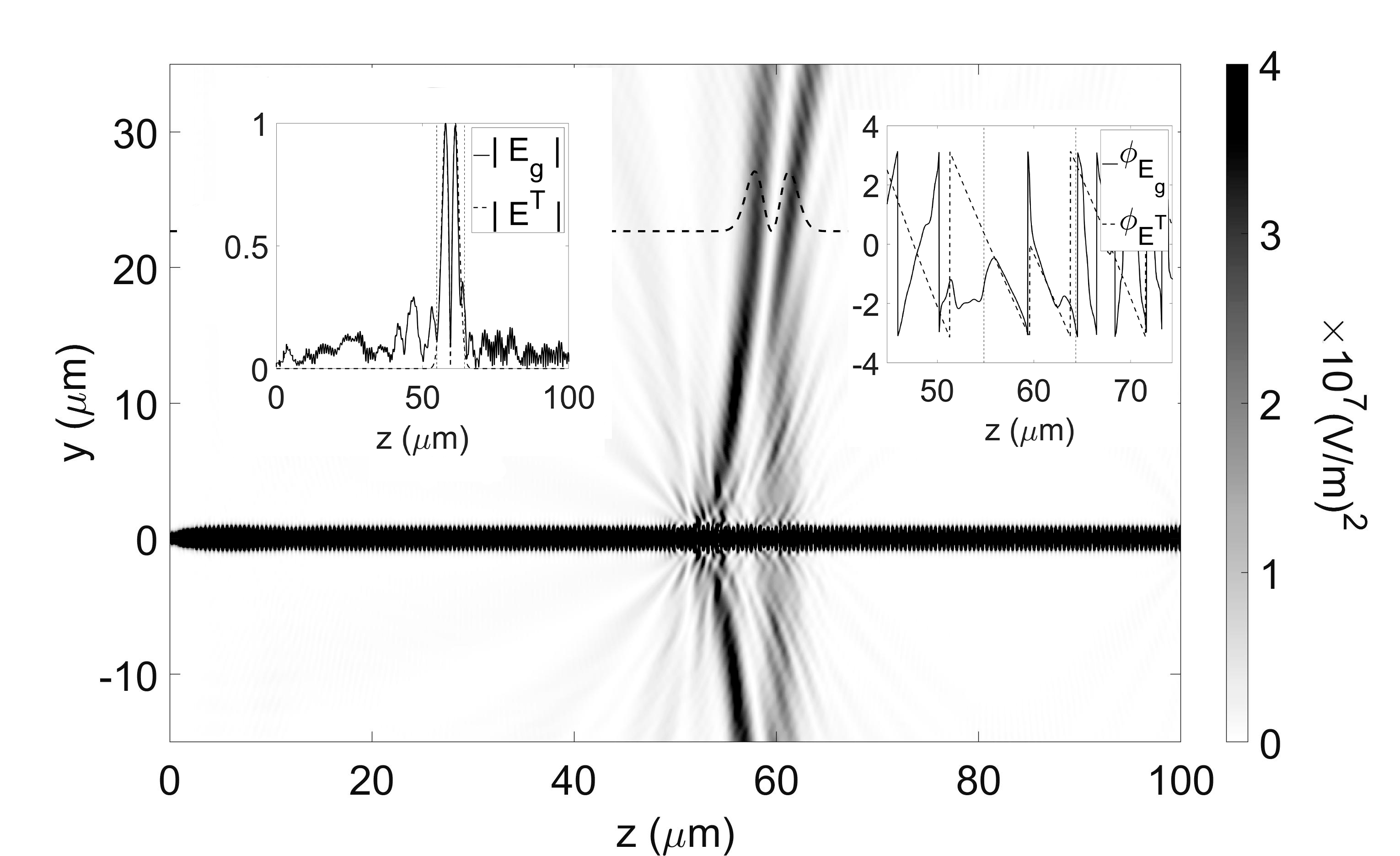}}\\
	\subfloat[]{\includegraphics[width=7.7cm]{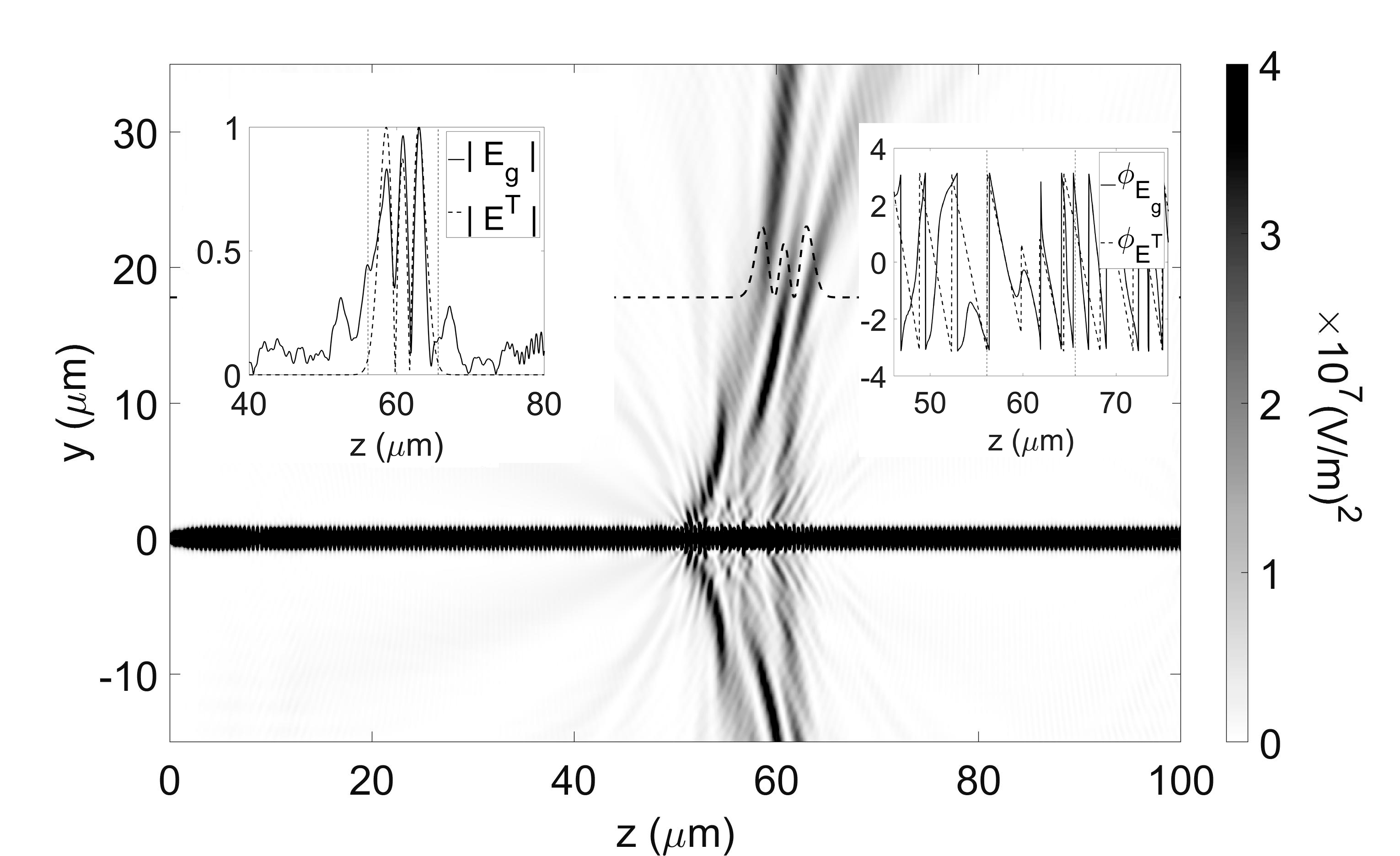}}
	\subfloat[]{\includegraphics[width=7.7cm]{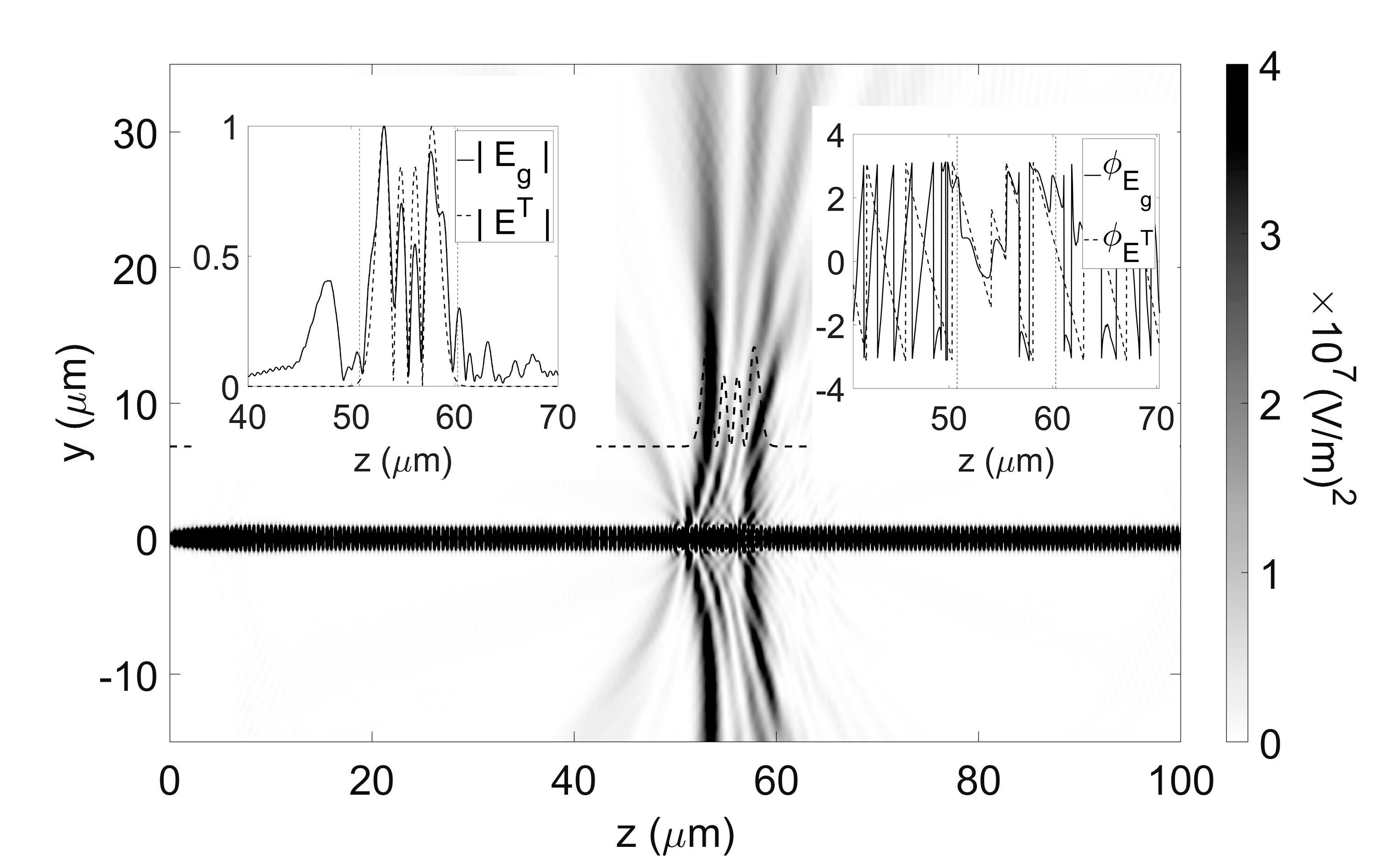}}
	\caption{Scattering maps for gratings optimized to couple light in a given mode of a multimode fiber. The dark horizontal line refers to the optical mode which propagates in the waveguide. The gray scale refers to the field intensity and is given on the right of the panel. The dashed lines refer to the optimum overlap profile at the best height. The insets show the normalized field amplitude (left) and the field phase (right) on the optimized \textit{y}, compared with the same quantities for the target fiber mode (dotted line). The vertical lines delimit the fiber core region. (a) grating to excite the HG00 mode, (b) grating to excite the HG10 mode, (c) grating to excite the HG20 mode, (d) grating to excite the HG30 mode.}
	\label{multiplexing}
\end{figure*}

The results are reported in Table \ref{valori} for N=25. Different grating geometries and different best positions for the fiber were found. A minimum coupling efficiency of 33 \% has been found. It is worth noticing that best coupling efficiencies are obtained when both the field amplitudes and the phases match on the input fiber facet (see Fig.\ref{multiplexing}).

The coupling selectivity can be estimated by calculating the efficiency to couple to the first order mode HG00 by the optimum grating and position, as shown in the last row of Table \ref{valori}. It is apparent that when the grating is optimized for a different order mode, the excitation of HG00 is strongly depressed.

\begin{table}[h!]
	\centering
	\caption{Performances of the optimum grating to couple in a given mode. N=40}
\label{valori}
	\begin{tabular}{cccccc}
		\toprule
		Parameter & Monomode fiber & HG00 & HG10 & HG20 & HG30\\
		\midrule
R & 1\% & 1\% & 1\% &0.9\%& 0.5\%\\
T & 0.2\% & 0.5\% & 1\%&1\%&0.4\%\\
d ($\mu$m) & 5.99 &5.32 & 9.55 & 10.77&5.48 \\
y ($\mu$m)& 34.99&26.61 & 22.67 & 17.79&6.80\\
$\alpha$ ($^0$)& 1.46 & 1.42 & 7.29 & 15.06& 11.88\\ 
$\Gamma$& 95\% &94\% &75\%& 68\%& 69\%\\
$\eta_c$& 47\% &46\% &37\%& 33\% & 34\%\\
$\eta_c$ (to HG00)& - &46\% & 0.03\%& 4.1\% & 0.2\%\\
		\bottomrule
	\end{tabular}
\end{table}

\section{Discussion of the limits of the proposed method}

The method is based on the assumption that only the waveguide perturbation (i.e. the SiO$_2$ layers) affects the scattering profile. The presence of other interfaces, such as the one at the surface or at the silicon substrate, is not considered. In addition, independence of the scattering events has been assumed. These two assumptions are satisfied as long as the refraction and reflection at these interfaces are negligible, and as long as the reflected light does not couple back in the grating. 

\begin{figure}[h!]
	\centering
	\subfloat[\label{CA}]{\includegraphics[scale=0.1]{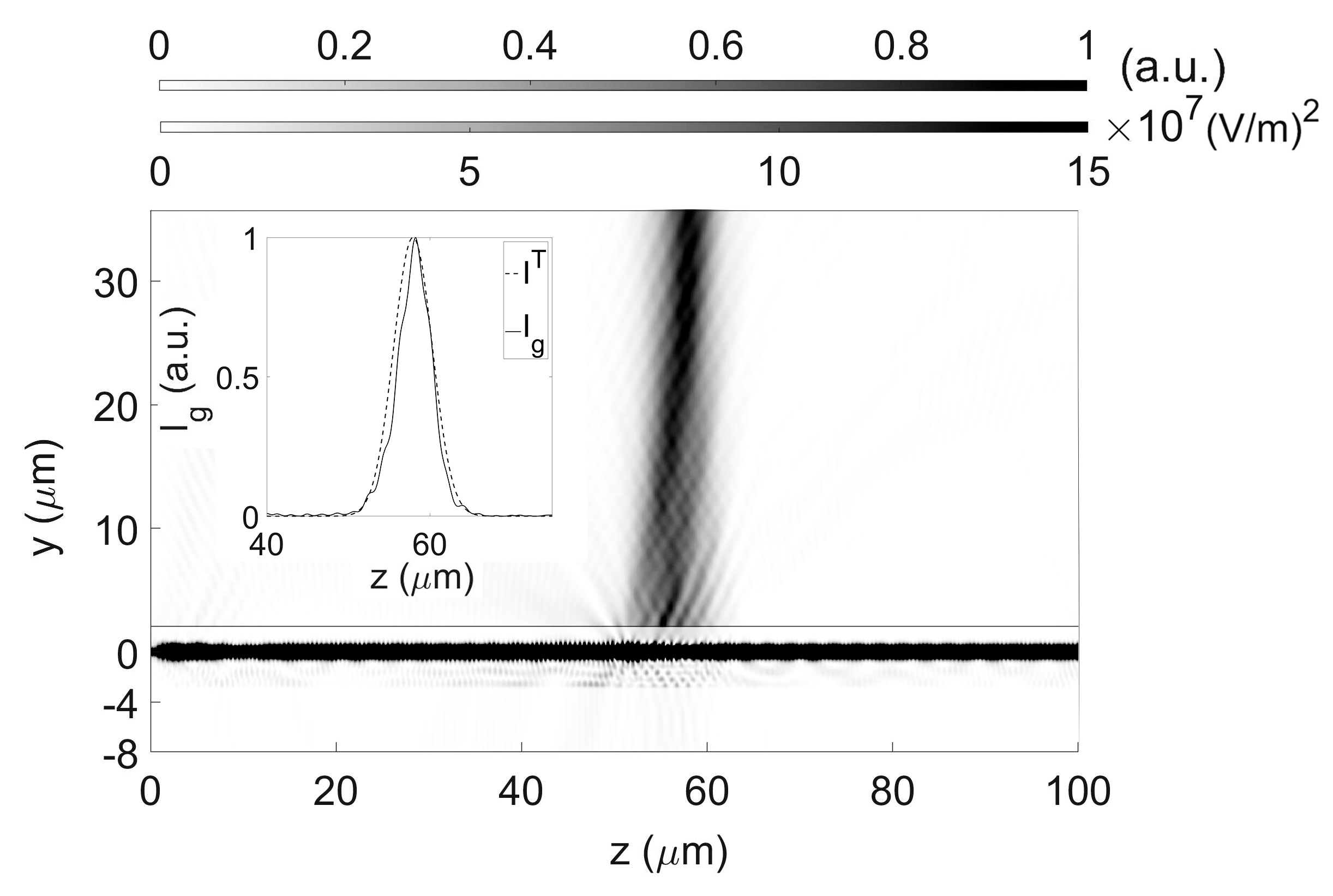}}\\
	\subfloat[\label{CA}]{\includegraphics[scale=0.1]{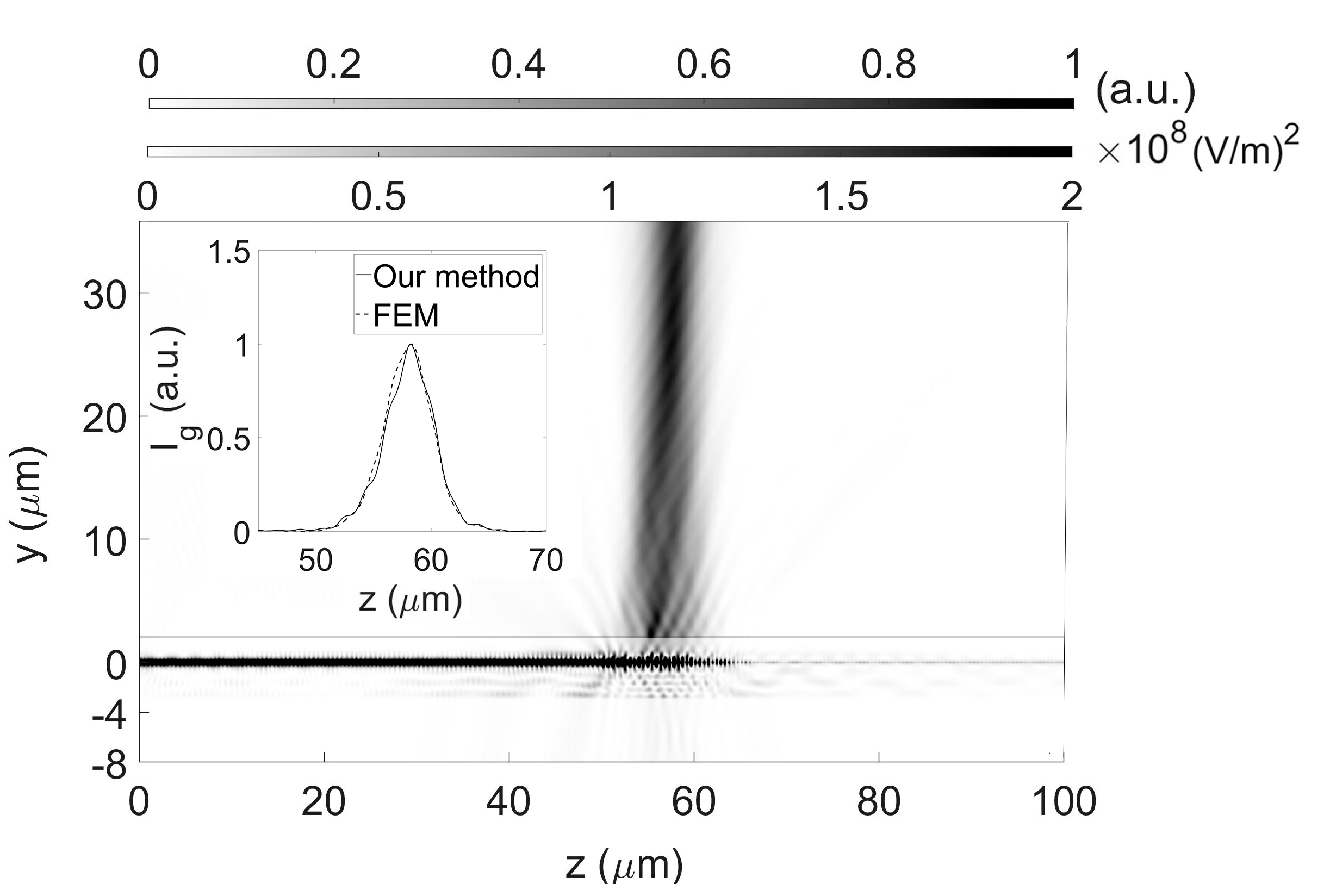}}\\
	\caption{Scattering maps simulated with (a) our method (b) FEM. In order to compare the 2D scattering maps, the images were splitted in two parts: for $y>$2 $\mu$m the scattered field is normalized (top color bar), while the lower part presents the not normalized field (bottom color bar). The inset in (a) shows the comparison between the 1D profile of the scattered field at the optimized $y$ and the target profile; the inset in (b) shows the comparison between the distributions at the optimized $y$ obtained with our method and FEM, from the same optimized grating geometry.}
	\label{ComsolAria}
\end{figure}

To validate the method, we performed an optimization of a grating to couple light in a single mode fiber with parameters extracted from a FEM simulation with both air and the silicon buffer in the simulation domain. Specifically, we assumed a 60 nm thick Si waveguide with 800 nm thick SiO$_2$ top cladding and 3 $\mu$m thick SiO$_2$ BOX (buried oxide) on a 5 $\mu$m thick Si substrate. The substrate reflects the down scattered light to the top cladding (comparing the pointing vectors at different y's equally distant from the waveguide, we found that 57\% of the light is scattered to the top). The optimization yields a grating with $\Gamma$=94\% and $\eta_c$=52.5\% (Fig. \ref{ComsolAria}a). We compared our solution to the same grating computed by FEM simulation (Fig. \ref{ComsolAria}b). The FEM simulated scattering profile at y=34.99 $\mu$m shows $\Gamma$=93 \%. These results show the accuracy of our method (e.g., few \% in the coupling efficiency estimate) and that the presence of other interfaces does not undermine the assumptions we used. In addition, it can be also noted that the optical mode transmitted by the grating is not correctly described by our method. In fact the method is suited for the scattering map calculations but not for the optical propagation mode estimate.

As a further test, we calculated the coupling efficiency of the optimized grating in the case of a random error on the layer lengths of $\pm$ 5 nm ($\pm$ 10 nm) to simulate the effects of the fabrication tolerance. In this case, $\Gamma$= 94 \% (92\%) and $\eta_c$= 52.5 \% (50 \%). It can be concluded that the obtained optimal grating sequence is not dependent on the accuracy of the single layer lengths. Therefore, the efficiency of the grating is preserved within reasonable fabrication tolerances.

\begin{figure}[h!]
	\centering
	\subfloat[\label{shall_draw}]{\includegraphics[scale=0.16]{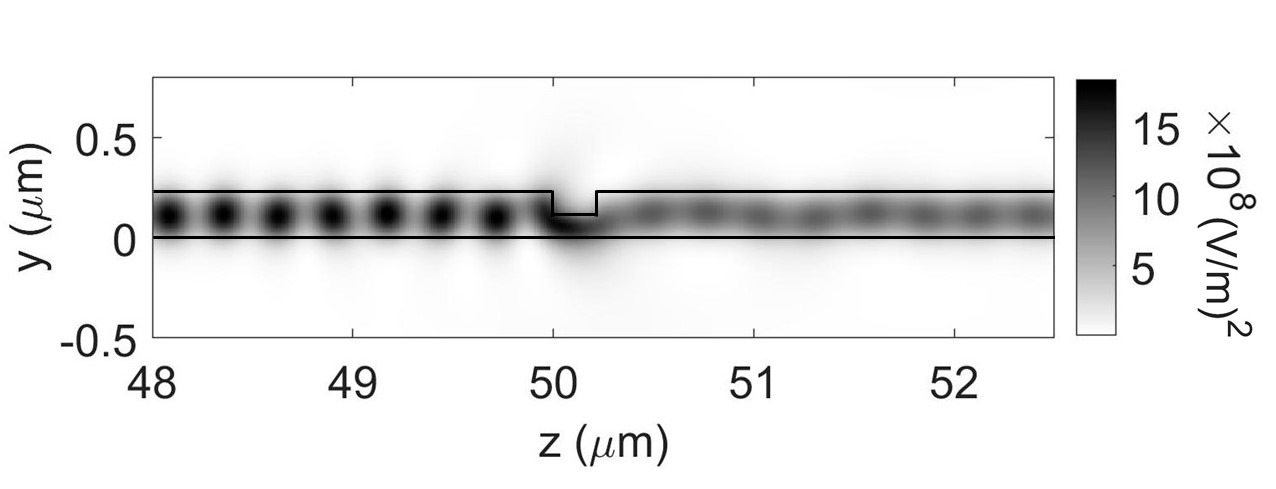}}\\
	\subfloat[\label{full_draw}]{\includegraphics[scale=0.16]{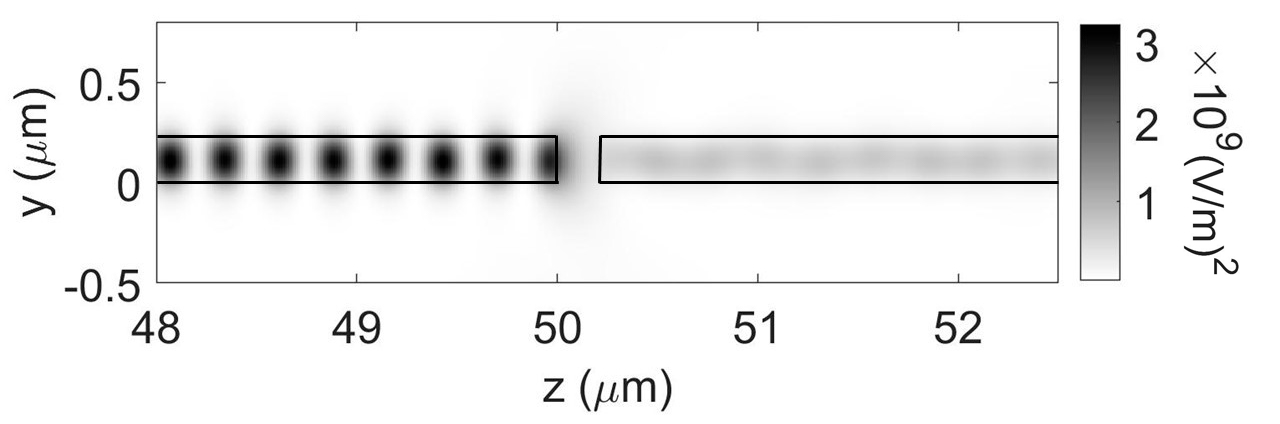}}
	\caption{FEM simulations of the fundamental mode propagating in (a) a shallow etched Si waveguide and in (b) a totally etched Si waveguide. The waveguide is 220 nm thick and the SiO$_2$ layer is 200 nm long. In (a) the Si is partially etched by 100 nm. The light wavelength is 1550 nm.}
	\label{shallowcoms}
\end{figure}

We also investigate the assumption of block independence in two situations: a grating with a totally etched waveguide and a grating with a shallow etched waveguide. In this last situation, the optical mode propagates also in the thinned part of the waveguide.
This couples the propagating fields before and after the SiO$_2$ layer, which, in turn, couples adjacent or far scattering events breaking the independence assumption. An example of this can be observed in Fig. \ref{shall_draw}, where a partially etched SiO$_2$ layer is inserted across a Si waveguide. The field propagating after the SiO$_2$ layer shows the typical undulation due to mode beating: this is due to the coupling of the mode propagating in the shallow etched section to the first and second mode of the following waveguide. The beating is then due to the coupling to the second mode. In addition to this, there is the interference between the forward and the backward propagating modes caused by the reflections, transmissions and scattering in the SiO$_2$ layer region. Note that in the case of a totally etched waveguide, the mode after the SiO$_2$ layer shows mostly a propagation character with negligible amplitude oscillations (Fig. \ref{full_draw}).

On the other hand, the totally etched geometry limits the efficiency of the system to about 50\%. One pragmatic way to overpass the modal coupling due to the partial waveguide etch is to monitor the degree of coupling as a function of the SiO$_2$ layer length for a given waveguide etch depth. For example, we found that for a 100 nm etching depth in a 220 nm thick Si waveguide the minimum SiO$_2$ layer length to preserve the block independence is of about 450 nm.

\section{Conclusions}

The FEM enhanced TM method proposed in this paper has been validated by few examples. The main advantages of the method are the reduced computation time, its customization to the different applications by properly defining the cost function and the ease implementation on small computers.  Validation of the design of the various grating has shown that the accuracy of the method is within few percent with respect to more time consuming and computational demanding FEM or FDTD methods. Extension to 3D simulations can be performed along the lines suggested in section \ref{sezione_neurone}, where the convolution between the optimized scattered distribution and waveguide propagation mode was performed.

A delicate issue is about the design of gratings where common practices to increase the grating efficiency, such as shallow etched gratings or holographic gratings or metal enhanced gratings \cite{Dirk:coup, bridging}, are used. The method presented in this paper cannot treat these systems. Specific extensions are required, e.g. by three dimensional modeling or by considering the multiple interference due to the reflections at the metal interface. However, in this paper we have preferred to limit ourselves to a simple and fast technique which allows designing gratings that can be easily and cheaply fabricated by most of the fabrication facilities.

\section*{Acknowledgments}
This project has received funding from the European Research Council (ERC) under the
European Union’s Horizon 2020 research and innovation programme (grant agreement No
788793, BACKUP), and from the MIUR under the project PRIN PELM (20177 PSCKT).

\bibliographystyle{ieeetr.bst}

\bibliography{A_FEM_enhanced_method}

\begin{thebibliography}{10}

\bibitem{Quaranta:resonant}
G.~Quaranta, G.~Basset, O.~J. Martin, and B.~Gallinet, ``Recent advances in
  resonant waveguide gratings,'' {\em Laser \& Photon. Rev.}, vol.~12, no.~9,
  p.~1800017, 2018.

\bibitem{McLamb:extr}
M.~McLamb, Y.~Li, S.~Park, M.~Lata, and T.~Hofmann, ``Diffraction gratings for
  uniform light extraction from light guides,'' in {\em IEEE 16th Int. Conf. on
  Smart Cities: Improving Quality of Life Using ICT \& IoT and AI (HONET-ICT)},
  pp.~220--222, 2019.

\bibitem{Ziel:planar}
I.~Zielonka and P.~Karasi{\'n}ski, ``Planar waveguides with diffracion gratings
  (rewiev paper),'' {\em Molecular and Quantum Acoustics}, vol.~22,
  pp.~293--302, 2001.

\bibitem{Silberstain:int}
E.~Silberstein, P.~Lalanne, J.-P. Hugonin, and Q.~Cao, ``Use of grating
  theories in integrated optics,'' {\em JOSA A}, vol.~18, no.~11,
  pp.~2865--2875, 2001.

\bibitem{khalid:apo}
K.~Khalid, M.~Zafrullah, S.~Bilal, and M.~Mirza, ``Simulation and analysis of
  gaussian apodized fiber bragg grating strain sensor,'' {\em J. Opt.
  Technol.}, vol.~79, no.~10, pp.~667--673, 2012.

\bibitem{ML:capo}
K.~Yao, R.~Unni, and Y.~Zheng, ``Intelligent nanophotonics: merging photonics
  and artificial intelligence at the nanoscale,'' {\em Nanophotonics}, vol.~8,
  no.~3, pp.~339--366, 2019.

\bibitem{Chen:coup}
X.~Chen, C.~Li, C.~K. Fung, S.~M. Lo, and H.~K. Tsang, ``Apodized waveguide
  grating couplers for efficient coupling to optical fibers,'' {\em IEEE
  Photon. Technol. Lett.}, vol.~22, no.~15, pp.~1156--1158, 2010.

\bibitem{Patri:coup}
A.~Patri, X.~Jia, M.~Mohsin, S.~K{\'e}na-Cohen, and C.~Caloz, ``Compact grating
  coupler using asymmetric waveguide scatterers,'' in {\em Frontiers in
  Optics}, pp.~JW4A--80, Optical Society of America, 2019.

\bibitem{Nambiar:coup}
S.~Nambiar, P.~Sethi, and S.~K. Selvaraja, ``Grating-assisted fiber to chip
  coupling for soi photonic circuits,'' {\em Appl. Sci.}, vol.~8, no.~7,
  p.~1142, 2018.

\bibitem{Tang:coup}
Y.~Tang, Z.~Wang, L.~Wosinski, U.~Westergren, and S.~He, ``Highly efficient
  nonuniform grating coupler for silicon-on-insulator nanophotonic circuits,''
  {\em Opt. Lett.}, vol.~35, no.~8, pp.~1290--1292, 2010.

\bibitem{Dirk:coup}
D.~Taillaert {\em et~al.}, ``Grating couplers for coupling between optical
  fibers and nanophotonic waveguides,'' {\em Jpn. J. Appl. Phys.}, vol.~45,
  no.~8R, p.~6071, 2006.

\bibitem{bridging}
W.~S. Zaoui {\em et~al.}, ``Bridging the gap between optical fibers and silicon
  photonic integrated circuits,'' {\em Opt. Exp.}, vol.~22, no.~2,
  pp.~1277--1286, 2014.

\bibitem{ML:mapping}
D.~Melati {\em et~al.}, ``Mapping the global design space of nanophotonic
  components using machine learning pattern recognition,'' {\em Nature
  Commun.}, vol.~10, no.~1, pp.~1--9, 2019.

\bibitem{tempi:coup}
J.~Covey and R.~T. Chen, ``Efficient perfectly vertical fiber-to-chip grating
  coupler for silicon horizontal multiple slot waveguides,'' {\em Opt.
  Express}, vol.~21, no.~9, pp.~10886--10896, 2013.

\bibitem{scattering_boundaries}
W.~Frei, ``Using perfectly matched layers and scattering boundary conditions
  for wave electromagnetics problems.'' \textit{COMSOL Blog},
  https://www.comsol.com/blogs/using-perfectly-matched-layers-and-scattering-boundary-conditions-for-wave-electromagnetics-problems/
  (accessed {Feb}. 23, 2021).

\bibitem{tempi:coup1}
T.~Grosges, A.~Vial, and D.~Barchiesi, ``Models of near-field spectroscopic
  studies: comparison between finite-element and finite-difference methods,''
  {\em Opt. Express}, vol.~13, no.~21, pp.~8483--8497, 2005.

\bibitem{tempi:camfr1}
P.~Bienstman and R.~Baets, ``Optical modelling of photonic crystals and vcsels
  using eigenmode expansion and perfectly matched layers,'' {\em Opt. Quantum
  Electron.}, vol.~33, no.~4, pp.~327--341, 2001.

\bibitem{tempi:camfr2}
J.~{\v{C}}tyrok{\`y} {\em et~al.}, ``Bragg waveguide grating as a 1d photonic
  band gap structure: Cost 268 modelling task,'' {\em Opt. Quantum Electron.},
  vol.~34, no.~5, pp.~455--470, 2002.

\bibitem{part_swarm}
J.~C. Bansal, P.~K. Singh, and N.~R. Pal, {\em Evolutionary and swarm
  intelligence algorithms}.
\newblock Springer, 2019.

\bibitem{part_swarm1}
Y.~Ma {\em et~al.}, ``Ultralow loss single layer submicron silicon waveguide
  crossing for soi optical interconnect,'' {\em Opt. Express}, vol.~21, no.~24,
  pp.~29374--29382, 2013.

\bibitem{part_swarm2}
T.~Watanabe, M.~Ayata, U.~Koch, Y.~Fedoryshyn, and J.~Leuthold, ``Perpendicular
  grating coupler based on a blazed antiback-reflection structure,'' {\em J.
  Light. Technol.}, vol.~35, no.~21, pp.~4663--4669, 2017.

\bibitem{welkenhuysen2016integrated}
M.~Welkenhuysen {\em et~al.}, ``An integrated multi-electrode-optrode array for
  in vitro optogenetics,'' {\em Sci. Rep.}, vol.~6, no.~1, pp.~1--10, 2016.

\bibitem{Chang}
W.~S. Chang, {\em \textit{Fundamentals of guided-wave optoelectronic devices}}.
\newblock Cambridge University Press, 2009.

\end{thebibliography}

\vfill
\begin{IEEEbiographynophoto}{Clara Zaccaria}
received the master degree in physics in 2018 from the University of Padua. She is currently working toward the Ph.D. degree at the University of Trento. Her current research interests include integrated photonics and biophysics.
\end{IEEEbiographynophoto}
\vfill
\begin{IEEEbiographynophoto}{Mattia Mancinelli}
received the Ph.D. degree in physics in 2013 from the University of Trento, Trento, Italy. After 4 years of post-doctoral research in the field of integrated optics and non-linear optics and 2 years of R$\&$D in the telecom company SMoptics he is currently a tenure track researcher at University of Trento in the field of neuromorphic photonics. He is currently involved in the ERC BACKUP project and ERC-PoC ALPI. He is author and co-author of more than 30 papers and he has supervised 7 master students.
\end{IEEEbiographynophoto}

\vfill
\begin{IEEEbiographynophoto}{Lorenzo Pavesi}
(M’08–SM’11–F’16) received the Ph.D. degree in physics from the Ecole Polytechnique Federale of Lausanne, Lausanne, Switzerland, in 1990. He is currently a Full Professor in experimental physics and the Head of the Nanoscience Laboratory, Department of Physics, University of Trento, Trento, Italy. He has authored or co-authored more than 400 papers, authored several reviews and two books, and edited more than 10 books. He holds ten patents. His current research interests include silicon photonics, integrated quantum photonics, and neuromorphic photonics. He was a recipient of the title of Cavaliere by the Italian President for scientific merit in 2001. In 2010 and 2011, he was elected as a Distinguished Speaker of the IEEE-Photonics Society. He was the First President and the Founder of the IEEE Italian Chapter on Nanotechnology. He is an ERC grantee. He is Fellow of the SPIE and SIF.
\end{IEEEbiographynophoto}
\vfill

\vfill

\enlargethispage{-5in}

\end{document}